\documentclass[fleqn,usenatbib]{mnras}

\usepackage{newtxtext,newtxmath}

\usepackage[T1]{fontenc}

\DeclareRobustCommand{\VAN}[3]{#2}
\let\VANthebibliography\thebibliography
\def\thebibliography{\DeclareRobustCommand{\VAN}[3]{##3}\VANthebibliography}

\usepackage{graphicx}	
\usepackage{amsmath}	
\usepackage{orcidlink}
\usepackage{multicol}
\usepackage{makecell}

\title[Dense gas tracers in and between spiral arms]{Dense gas tracers in and between spiral arms: from Giant Molecular Filaments to star-forming clumps}

\author[O. Feh\'er et al.]{
Orsolya Feh\'er,$^{1}$\thanks{E-mail: fehero@cardiff.ac.uk}\orcidlink{0000-0002-0786-7307}
S. E. Ragan$^{1}$\orcidlink{0000-0003-4164-5588}
F. D. Priestley$^{1}$\orcidlink{0000-0002-5858-6265}
P. C. Clark$^{1}$\orcidlink{0000-0002-4834-043X}
\\
$^{1}$School of Physics and Astronomy, Cardiff University, Queen's Buildings, The Parade, Cardiff C24 3AA, UK
}

\date{Accepted XXX. Received YYY; in original form ZZZ}

\pubyear{\the\year{}}

\begin{document}
\label{firstpage}
\pagerange{\pageref{firstpage}--\pageref{lastpage}}
\maketitle

\begin{abstract}
Giant Molecular Filaments are opportune locations in our Galaxy to study the star-forming interstellar matter and its accumulation on spatial scales comparable to those now becoming available for external galaxies. We mapped the emission of HCN(1$-$0), HCO$^+$(1$-$0), and N$_2$H$^+$(1$-$0) towards two of these filaments, one associated with the Sagittarius arm and one with an interarm area. Using the data alongside the COHRS $^{12}$CO(3$-$2), the CHIMPS $^{13}$CO(3$-$2), and \textit{Herschel}-based column density maps, we evaluate the dense gas tracer emission characteristics and find that although its filling factor is the smallest among the studied species, N$_2$H$^+$ is the best at tracing the truly dense gas. Significant differences can be seen between the $^{13}$CO, HCN, and $N$(H$_2$)$_{\mathrm{dust}}$ levels of the arm and interarm, while the N$_2$H$^+$ emission is more uniform regardless of location, meaning that the observed variations in line ratios like N$_2$H$^+$/HCN or N$_2$H$^+$/$^{13}$CO are driven by species tracing moderate-density gas and not the star-forming gas. In many cases, greater variation in molecular emission and ratios exist between regions inside a filament than between the arm and interarm environments. The choice of measure of the dense gas and the available spatial resolution have deep impact on the multi-scale view of different environments inside a galaxy regarding molecular emissions, ratios, and thus the estimated star formation activity.
\end{abstract}

\begin{keywords}
ISM: molecules -- ISM: structure -- galaxies: star formation
\end{keywords}



\section{Introduction}

\label{sec:intro}

Galactic star formation is concentrated in filamentary molecular clouds, the physical properties of which determine the characteristics of star formation within \citep{andre2014, hacar2023}. It has been shown that in nearby clouds the amount of star formation scales with the fraction of gas above a density threshold \citep{lada2010, wu2010}, the so-called dense gas mass fraction, and for external galaxies, the surface density of gas and star formation activity are similarly related averaged over a galaxy disk \citep{schmidt1959, kennicutt1998, gao2004, schinnerer2024}. However, higher angular resolution surveys reveal significant galaxy-to-galaxy variations \citep{bigiel2008, shetty2014}, and whether the correlation of star formation with dense gas holds up in all Galactic environments is not well known either \citep{eden2013, ragan2016, schinnerer2017}. The interpretation of these observations is hindered by insufficient knowledge of tracers, incongruity in their usage, angular resolution limits in observing nearby galaxies, and the lack of well-sampled maps of dense gas tracers (DGTs) towards extended Galactic environments.

In our Galaxy, unbiased, large surveys of multiple isotopologues and transitions of CO \citep[e.g.][]{jackson2006,barnes2015, dempsey2013, rigby2016} and infrared dust \citep[e.g.][]{benjamin2003, molinari2010} enable us to systematically probe the environmental dependence of star formation. The largest velocity-coherent objects over tens to hundred parsec-scales, the so-called Giant Molecular Filaments (GMFs), are good candidates as their size is comparable to clouds we are now able to resolve in nearby galaxies \citep[e.g.][]{stuber2023}. After the first GMF, ``Nessie'' \citep{jackson2010}, systematic searches using infrared extinction and CO or DGT-based velocity information resulted in samples of both spiral arm and interarm GMFs \citep{ragan2014, abreu2016}. Further studies compiled different catalogues in a number of other ways, e.g. the ``bones'' \citep{zucker2015}, the \textit{Herschel} large filaments \citep{wang2015} and the Minimum Spanning Tree filaments \citep{wang2016}. Significant variation in the properties of GMFs between and within catalogues suggest the groups may have unique formation or evolution mechanisms \citep{zucker2018}. The deeper investigation of GMFs is an excellent opportunity to map the dense gas in large-scale environments to investigate our tracers, the biases in the parameters used to measure the dense gas fraction, and the environmental effects influencing Galactic star formation, on our way to confirming or disproving the existence of a ``global Kennicutt-Schmidt relation''.

N$_2$H$^+$(1$-$0) was recently found, both observationally and in theoretical models, to be an excellent tracer of dense material, perhaps even better than HCN \citep{kauffmann2017, priestley2023a, priestley2023b}. We recently investigated N$_2$H$^+$(1$-$0) emission versus $^{13}$CO(3$-$2) \citep[CHIMPS,][]{rigby2016} and column density \citep{marsh2017} in two new GMFs, one in the Sagittarius spiral arm and one in an interarm area \citep{feher2024}. The results emphasized the many scale-, environment-, chemistry-, and excitation-dependent factors influencing the parameters used in describing the dense gas fraction and the emission characteristics of our tracers. Recent Galactic studies presented results concerning the variation (or lack thereof) of the ratio N$_2$H$^+$/HCN regarding Galactic molecular cloud environments \citep{pety2017, barnes2020} and external galaxies \citep{jimenez2023, stuber2023} as well. Since HCN is now known as a poor tracer of higher than 10$^4$\,cm$^{-3}$ volume density gas \citep{evans2020, jones2023, priestley2024}, which is the star formation threshold proposed by \citet{lada2010}, the N$_2$H$^+$/HCN ratio may infer the star formation efficiency (SFE). Smaller-than-beamsize effects towards external galaxies in our current and future observations need to be investigated: it is still a question whether the emission seen in extragalactic observations originates from clouds similar to local environments.

In this study, we investigate the same two GMFs identified in the CHIMPS survey that have been mapped in DGT emission with the IRAM 30\,m telescope. Our previous paper was focused on the N$_2$H$^+$(1$-$0) emission \citep{feher2024}, now we add HCN(1$-$0), HCO$^+$(1$-$0) and analyze their emission characteristics alongside $^{12}$CO(3$-$2) from COHRS \citep{dempsey2013}, $^{13}$CO(3$-$2) from CHIMPS, and the H$_2$ column density based on the Hi-GAL survey \citep{molinari2010}. For DGTs in GMFs see also \citet{wang2020}. In Section~\ref{sec:data} we describe the datasets and the data reduction methods. In Section~\ref{sec:results} we first present the general emission characteristics of the observed DGTs, then derive molecular ratios, assess the results for the arm and the interarm, and compare both to recent studies of Galactic and extragalactic molecular clouds. From Section~\ref{sec:discuss} we consider the variation in the measured parameters from filament-scale to clump-scale and across density regimes and environments, then we summarize our findings in Section~\ref{sec:concl}.

\section{Observations and data reduction} 
\label{sec:data}

\subsection{The CHIMPS and COHRS surveys}

The two GMFs we targeted were identified in the position-velocity diagrams of the $^{13}$CO/C$^{18}$O (J\,=\,3$-$2) Heterodyne Inner Milky Way Plane Survey (CHIMPS) which was carried out on the 15\,m James Clerk Maxwell Telescope (JCMT) using the Heterodyne Array Receiver Program (HARP). The survey covers around 18 square degrees between 27.5\,$\leq l \leq$\,46.4$^{\circ}$ Galactic longitude and |b|\,$\leq$\,0.5$^{\circ}$ Galactic latitude. The angular resolution of the observations is 15\,$\arcsec$ and the velocity width per channel is 0.055\,km\,s$^{-1}$. The data is presented in units of $T_{\mathrm{A}}^*$ antenna temperatures which can be converted to main beam brightness temperature with $T_{\mathrm{MB}}$\,=\,$T_{\mathrm{A}}^*$/$\eta_{\mathrm{MB}}$ where $\eta_{\mathrm{MB}}$\,=\,0.72. The numbers reported in this paper are on the $T_{\mathrm{MB}}$ scale (in contrast with \citet{feher2024} where it remained in the $T_{\mathrm{A}}^*$ scale). For more information on the survey and the filaments see \citet{rigby2016} and \citet{feher2024}.

The two GMFs studied in this paper have no full counterparts in the previously discussed GMF surveys. The filament GMF38.1-32.4 by \citet{ragan2014} extends into the western regions of our interarm GMF (Region\,1 and 2, see later), however, the $v_{\mathrm{LSR}}$ values for that object are lower (50$-$60\,km\,s$^{-1}$) than what could be associated with our interarm cloud. The western part of our spiral arm GMF (part of Region\,5) appears in the filament catalogue of \citet{wang2016} as F37 at the velocities corresponding to the $v_{\mathrm{LSR}}$ of the arm GMF.

The $^{12}$CO(J\,=\,3$-$2) maps of the CO High-Resolution Survey (COHRS) are also included in our analysis. This survey was also carried out on the JCMT with the HARP instrument. The angular resolution at the frequency of the observed transition is 14$\arcsec$ and the provided velocity resolution is 0.42\,km\,s$^{-1}$. We downloaded the R2 data cubes provided in the CANFAR data archive which are in half-degree width spectral cubes in longitude, re-binned in velocity to 1\,km\,s$^{-1}$ channel widths. The intensities are expressed in units of $T_{\mathrm{A}}^*$ which can be converted to $T_{\mathrm{MB}}$ by dividing by $\eta_{\mathrm{MB}}$\,=\,0.61. The numbers reported in this paper are on the $T_{\mathrm{MB}}$ scale. For more information on the survey, see \citet{dempsey2013} and \citet{park2023}.

\subsection{The IRAM 30\,m on-the-fly maps}
\label{data:iram}

\begin{figure}
    \centering
    \includegraphics[width=\linewidth]{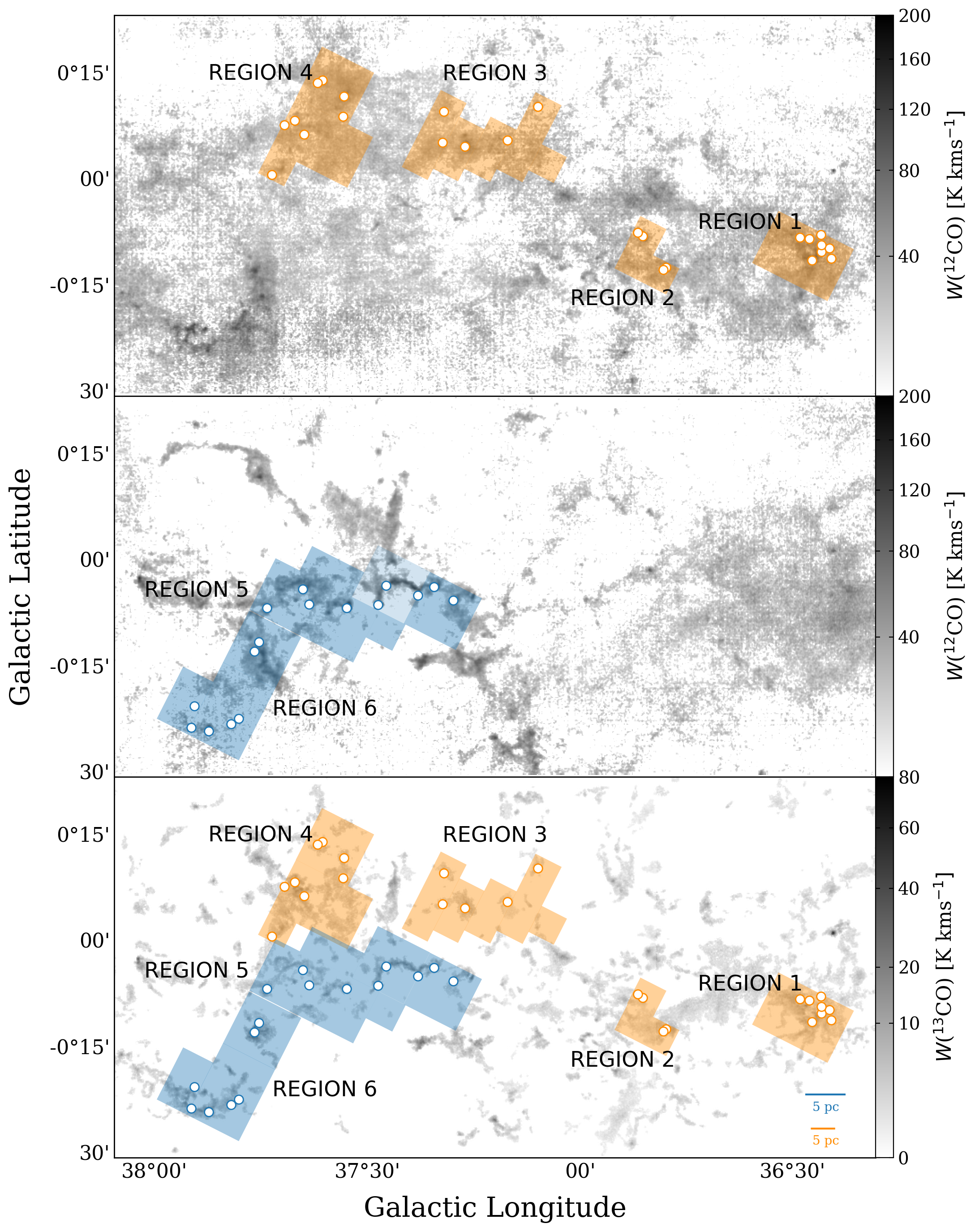}
    \caption{Overview of the two targeted filaments. The top and middle maps show the $^{12}$CO(3$-$2) integrated intensity maps from COHRS, first integrated in the 60$-$105\,km\,s$^{-1}$ then in the 25$-$61\,km\,s$^{-1}$ velocity interval. The bottom map shows the $^{13}$CO(3$-$2) integrated intensity map from CHIMPS integrated over all the velocity channels of the survey. Orange and blue rectangular areas mark the observing regions of the interarm and arm filaments, respectively, that were covered with the IRAM 30\,m telescope observations, their identifiers ranging from Region\,1 to 6. Circles mark the positions of the N$_2$H$^+$-clumps identified by \citet{feher2024}. The greyscale maps are normalized with a square-root function to better visually represent fainter structures.}
    \label{fig:overview}
\end{figure}

The observations of HCN(1$-$0), HCO$^+$(1$-$0) and N$_2$H$^+$(1$-$0) were performed with the IRAM 30\,m telescope during 2017 August 8$-$14 and 2023 February 22. The analysis of the N$_2$H$^+$(1$-$0) line was performed and published in \citet{feher2024}. The spectra of the HCN(1$-$0) and HCO$^+$(1$-$0) were obtained simultaneously to those observations, therefore they have the same spectral resolution of 200\,KHz, and a similar average rms noise of around 0.1\,K (in units of $T_{\mathrm{A}}$ antenna temperature). The half-power beam-width of the telescope at the tuning frequency of 91.3\,GHz was 26$\farcs$9, this means 26$\farcs$4 at the frequency of 93.173\,GHz for N$_2$H$^+$(1$-$0), 27$\farcs$6 at the frequency of 89.188\,GHz for HCO$^+$(1$-$0), and 27$\farcs$75 at the frequency of 88.631\,GHz for HCN(1$-$0). The reduction steps of correcting for the contaminated off-positions, conversion from $T_{\mathrm{A}}$ to $T_{\mathrm{MB}}$ main beam brightness temperature, baseline subtraction, extraction of target lines, then processing of the OTF maps to a common coordinate system and the imaging were done similarly to the reduction steps of N$_2$H$^+$(1$-$0). For detailed information on these see \citet{feher2024}. At the end of the reduction, spectral cubes of pixel sizes of 5$\farcs$55 and velocity resolutions of around 0.6\,kms$^{-1}$ in units of $T_{\mathrm{MB}}$ were made. Fig.~\ref{fig:overview} shows the coverage of the observations against the $^{12}$CO(3$-$2) and $^{13}$CO(3$-$2) maps of COHRS and CHIMPS for both filaments.

\subsection{Hi-GAL PPMAP continuum maps}

Similarly to the analysis by \citet{feher2024} we use the H$_2$ column density and dust temperature maps derived with PPMAP (point process mapping) by \citet{marsh2017} from the Herschel Infrared Galactic Plane Survey \citep[Hi-GAL,][]{molinari2010} maps in five infrared bands (70, 160, 250, 350, 500\,$\mu$m). Given a set of observational images of dust emission at multiple wavelengths, the associated point-spread functions, a dust opacity law, a grid of temperature values, and assuming optically thin dust, the PPMAP tool performs a non-hierarchical Bayesian process and outputs a cube of differential column densities as a function of angular position and dust temperature, a corresponding image cube of uncertainties and two more maps: $N$(H$_2$)$_{\mathrm{dust}}$ total column density and $T_{\mathrm{dust}}$ density-weighted mean dust temperature. We use these last two maps which have 12$\arcsec$ spatial resolution and pixel sizes of 6$\arcsec$. For more information on Hi-GAL, PPMAP, and the maps used in this paper see \citet{marsh2017, marsh2015} and \citet{feher2024}. 

\section{Results} 
\label{sec:results}

\subsection{Integrated intensity maps}

\begin{figure}
    \centering
    \includegraphics[width=0.95\linewidth]{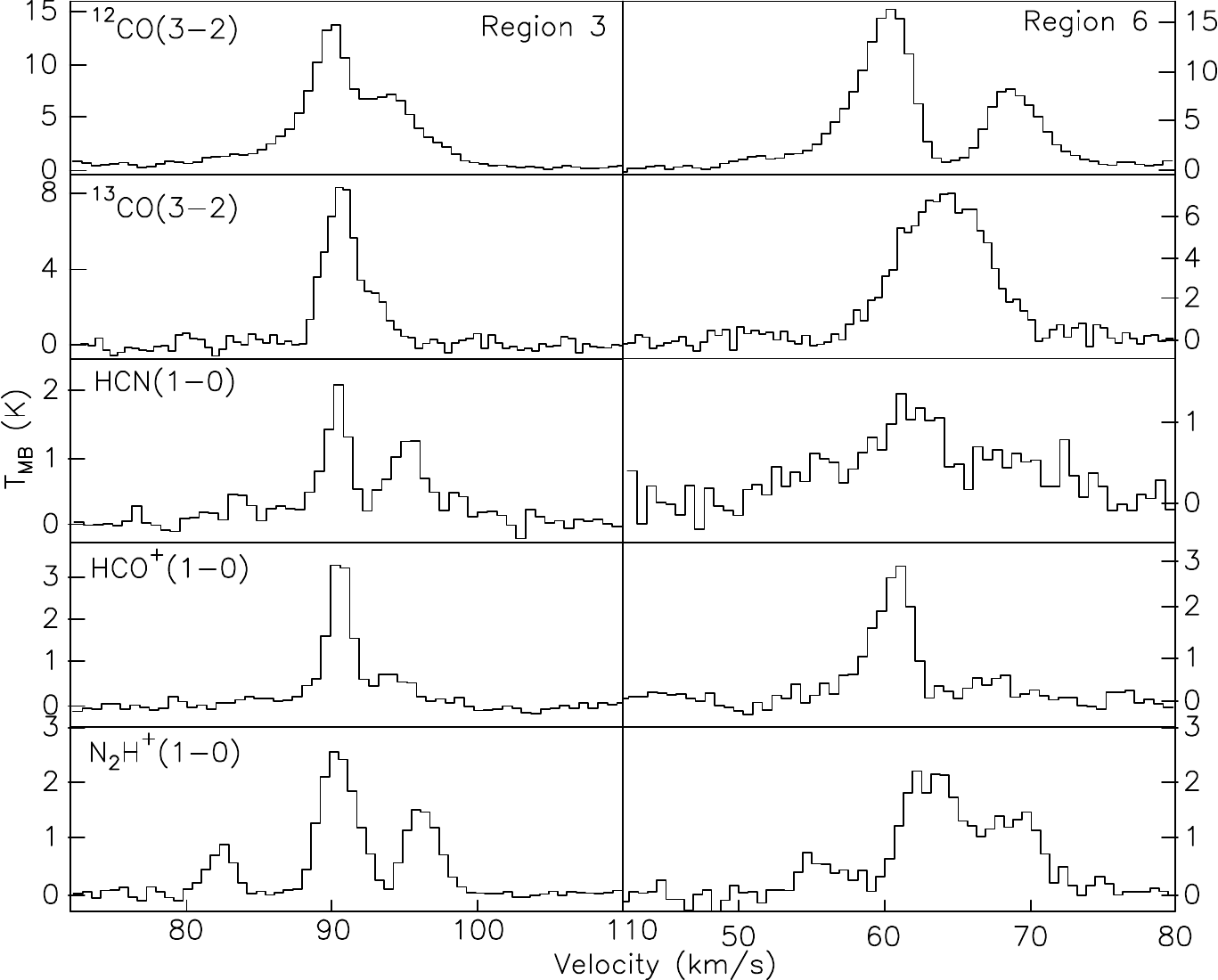}
    \caption{Example spectra of the studied species and transitions. The spectra are averaged in one IRAM 30\,m beam at the positions of the N$_2$H$^+$(1$-$0) integrated intensity maxima in the interarm and arm filament, in observing regions Region\,3 and Region\,6, respectively.}
    \label{fig:spectra}
\end{figure}

The HCN(1$-$0), HCO$^+$(1$-$0), and N$_2$H$^+$(1$-$0) lines were all detected towards each of our targets. Fig.~\ref{fig:spectra} shows example spectra of each investigated tracer. The hyperfine structure (HFS) of HCN(1$-$0) is usually blended at the line wings or even into a single line with two or three intensity peaks, while the HCO$^+$(1$-$0) line is usually well-approximated with a single Gaussian curve with smaller linewidths than HCN(1$-$0). As mentioned in \citet{feher2024}, the HFS components of N$_2$H$^+$(1$-$0) are generally blended into three groups or often into a single line with multiple peaks. The $v_{\mathrm{LSR}}$ central velocities of all three DGT lines were detected around the expected velocities of 50\,km\,s$^{-1}$ and 70\,km\,s$^{-1}$ associated with the Sagittarius spiral arm and the interarm structure identified on the CHIMPS position-velocity maps. We note that the HCO$^+$(1$-$0) line often shows more than these two velocity components in the 0$-$100\,km\,s$^{-1}$ interval.

\begin{figure*}
    \centering
    \includegraphics[width=\linewidth]{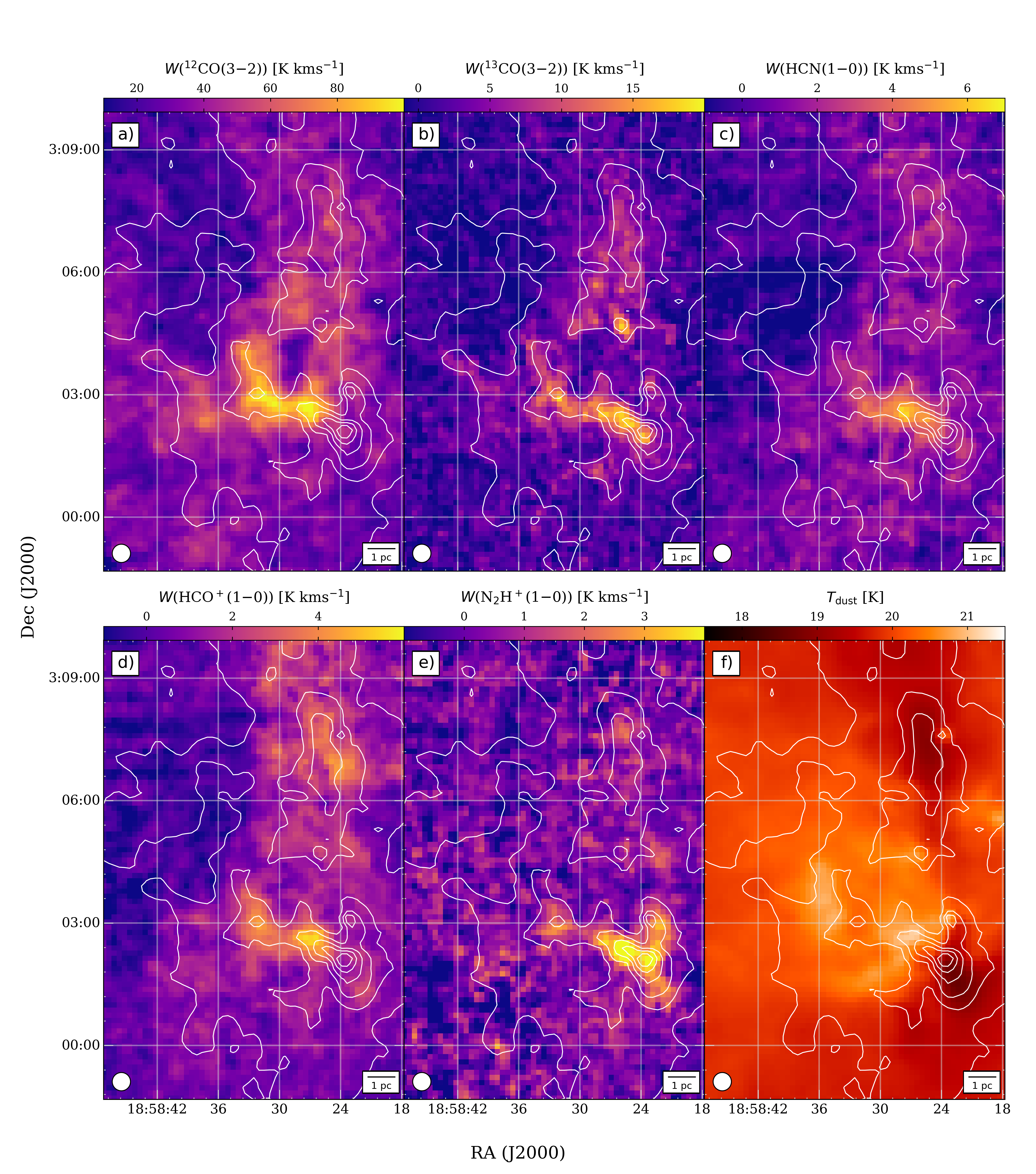}
    \caption{The a) $^{12}$CO(3$-$2), b) $^{13}$CO(3$-$2), c) HCN(1$-$0), d) HCO$^+$(1$-$0), e) N$_2$H$^+$(1$-$0) integrated intensity maps and f) the $T_{\mathrm{dust}}$ map of Region\,1. The white contours mark the $N$(H$_2$)$_{\mathrm{dust}}$ H$_2$ column densities at 1, 1.3...2.5\,$\times$\,10$^{22}$\,cm$^{-2}$. The primary beam size of the IRAM\,30m telescope and the spatial scale are both indicated on the maps.}
    \label{fig:reg1_morph}
\end{figure*}

\begin{figure*}
    \centering
    \includegraphics[width=\linewidth]{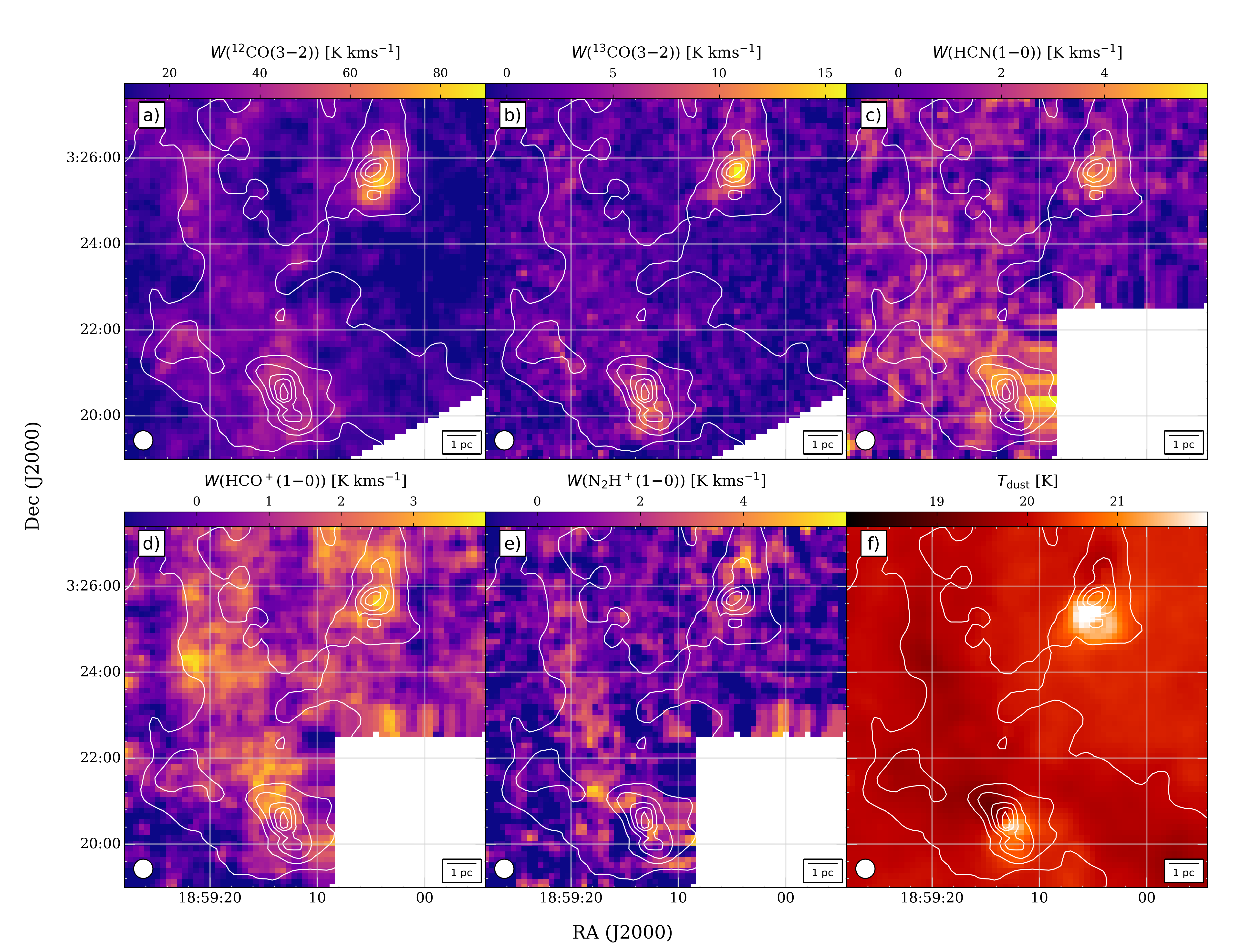}
    \caption{The same maps for Region\,2 as in Fig.~\ref{fig:reg1_morph}. The white contours mark the $N$(H$_2$)$_{\mathrm{dust}}$ H$_2$ column densities at 1, 1.25...2.5\,$\times$\,10$^{22}$\,cm$^{-2}$.}
    \label{fig:reg2_morph}
\end{figure*}

\begin{figure*}
    \centering
    \includegraphics[width=\linewidth]{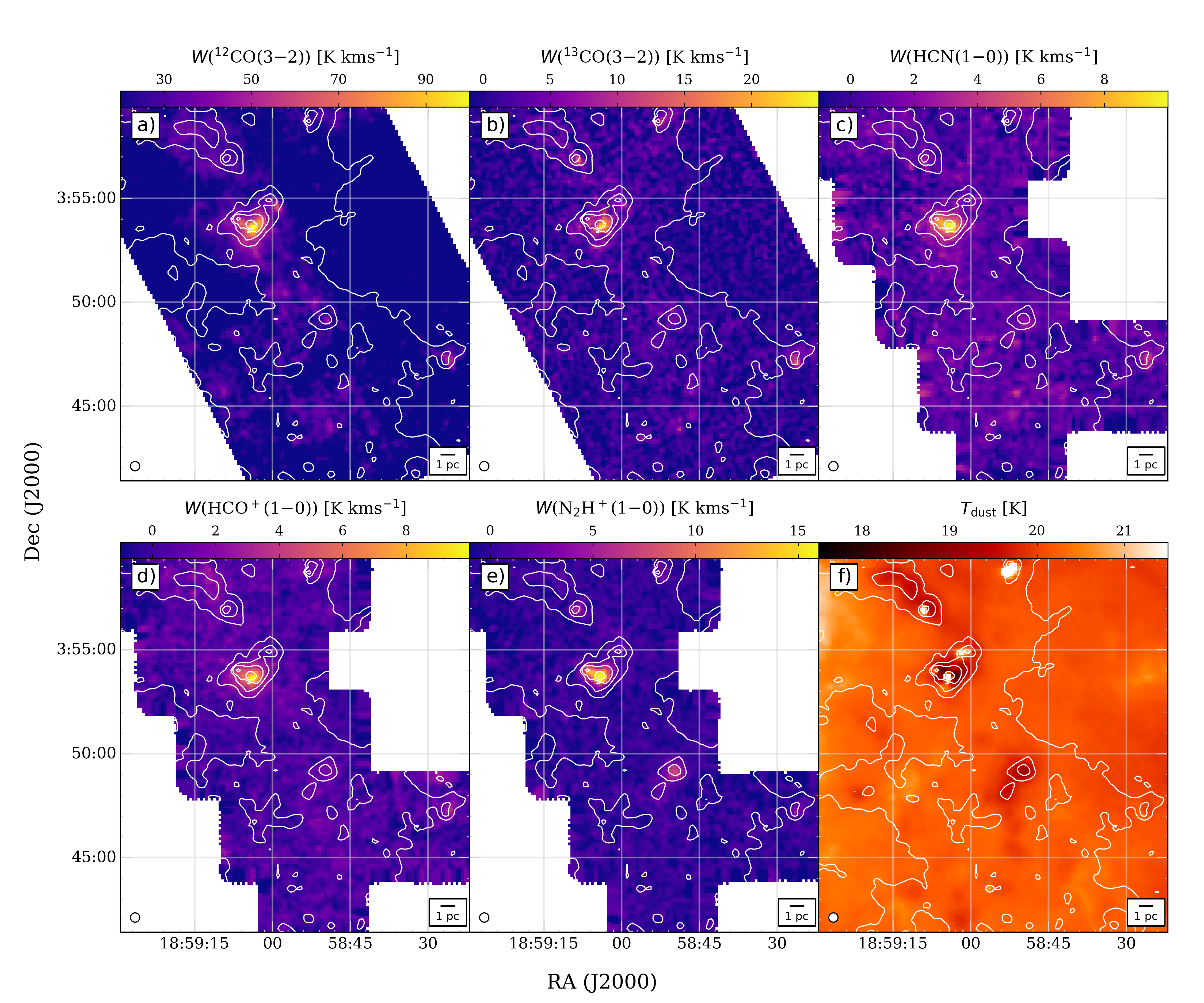}
    \caption{The same maps for Region\,3 as in Fig.~\ref{fig:reg1_morph}. The white contours mark the $N$(H$_2$)$_{\mathrm{dust}}$ H$_2$ column densities at 1, 1.25, 1.5, 2 and 4\,$\times$\,10$^{22}$\,cm$^{-2}$.}
    \label{fig:reg3_morph}
\end{figure*}

\begin{figure*}
    \centering
    \includegraphics[width=\linewidth]{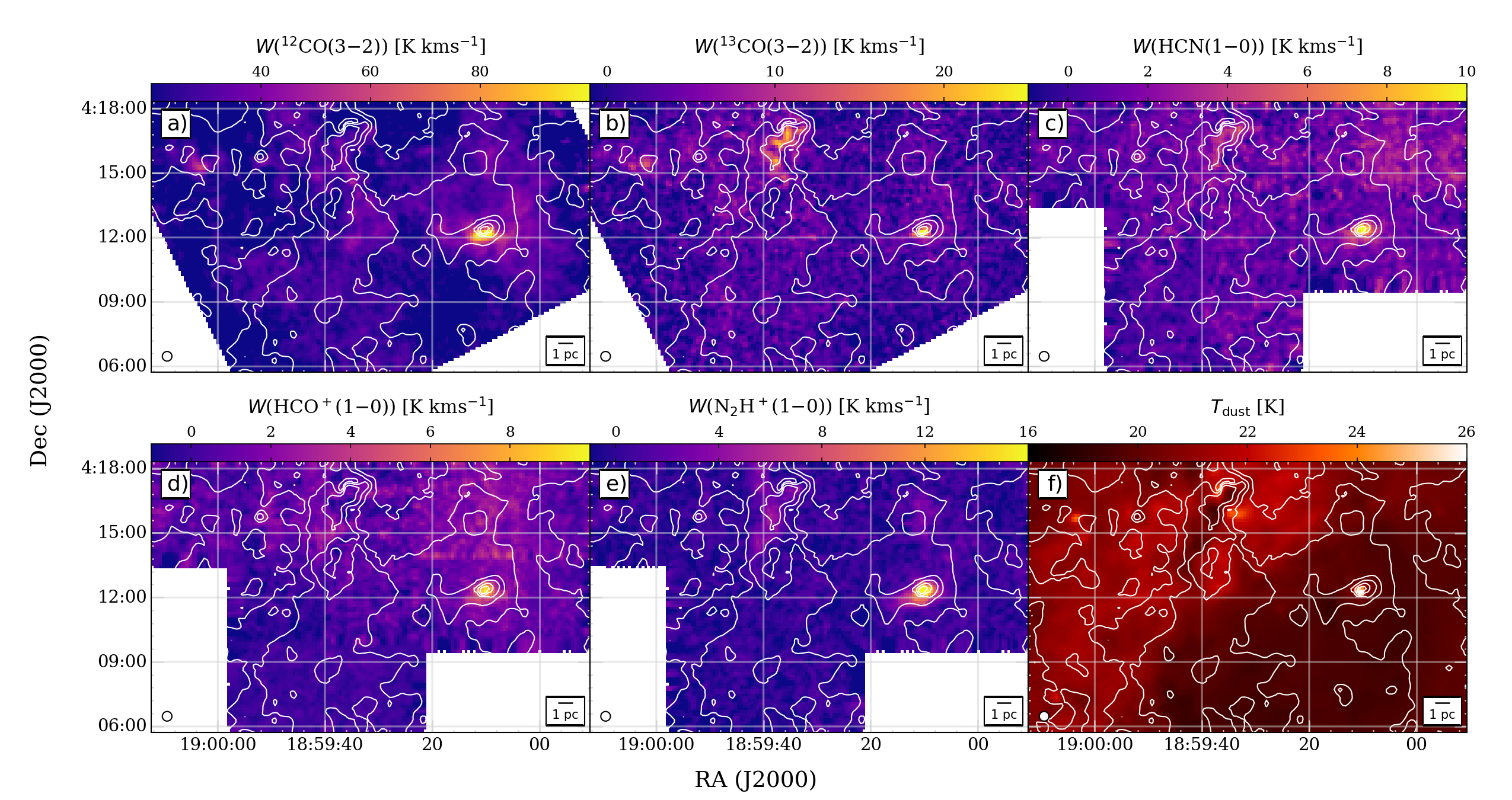}
    \caption{The same maps for Region\,4 as in Fig.~\ref{fig:reg1_morph}. The white contours mark the $N$(H$_2$)$_{\mathrm{dust}}$ H$_2$ column densities at 1, 1.25, 1.5, 2, 4 and 8\,$\times$\,10$^{22}$\,cm$^{-2}$.}
    \label{fig:reg4_morph}
\end{figure*}

\begin{figure*}
    \centering
    \includegraphics[width=0.9\linewidth]{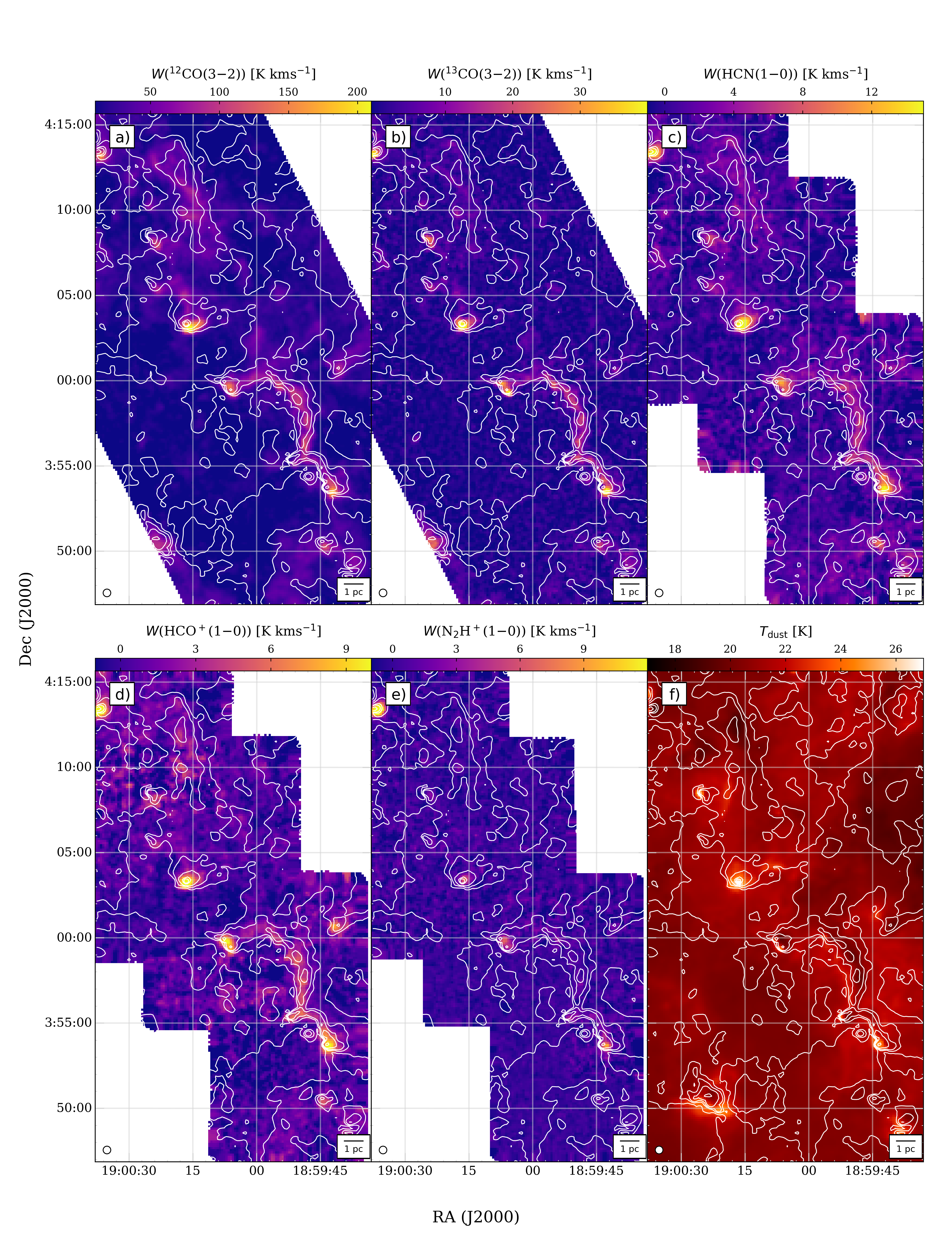}
    \caption{The same maps for Region\,5 as in Fig.~\ref{fig:reg1_morph}. The white contours mark the $N$(H$_2$)$_{\mathrm{dust}}$ H$_2$ column densities at 1, 1.25, 1.5, 2, 4, 6, and 12\,$\times$\,10$^{22}$\,cm$^{-2}$.}
    \label{fig:reg5_morph}
\end{figure*}

\begin{figure*}
    \centering
    \includegraphics[width=\linewidth]{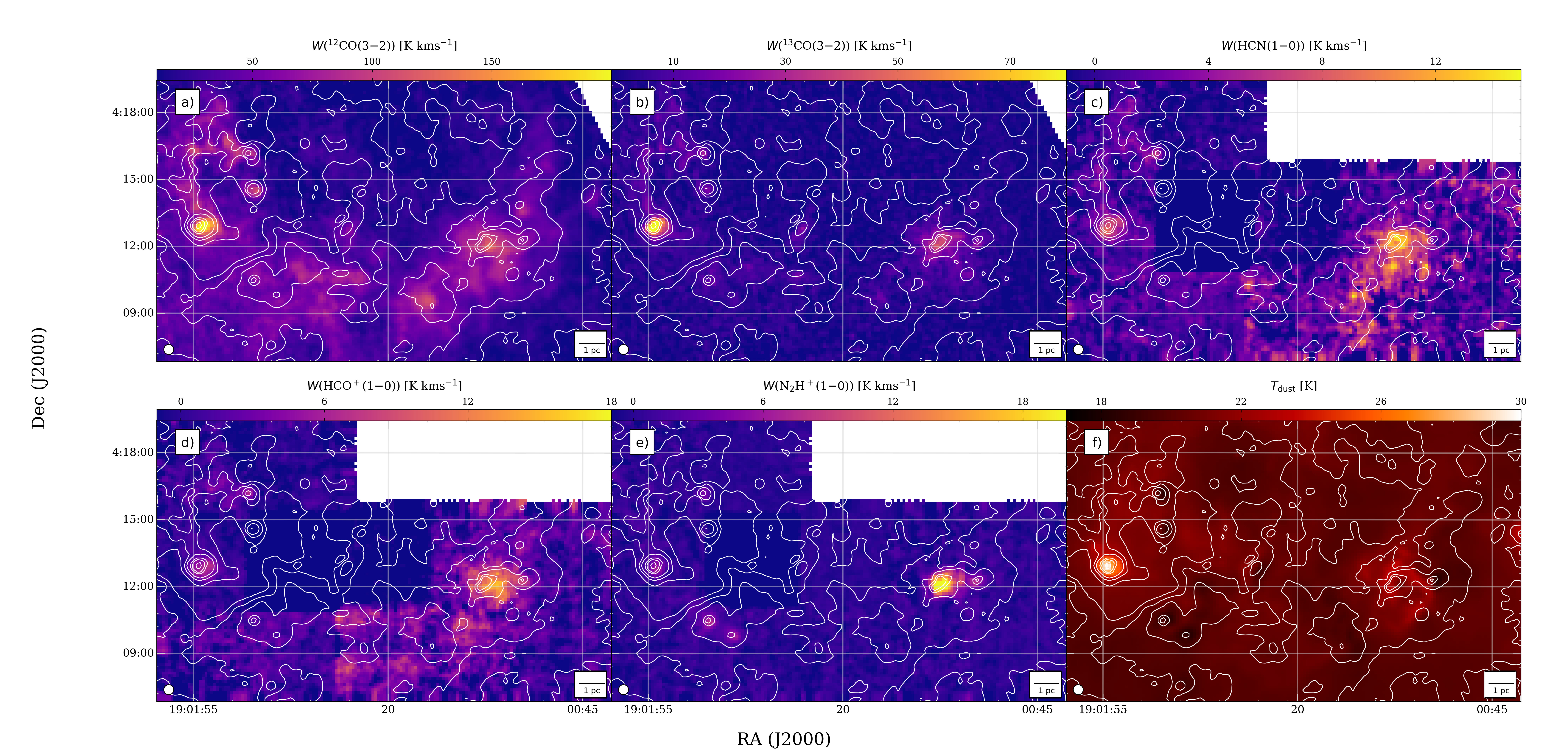}
    \caption{The same maps for Region\,6 as in Fig.~\ref{fig:reg1_morph}. The white contours mark the $N$(H$_2$)$_{\mathrm{dust}}$ H$_2$ column densities at 1, 1.25, 1.5, 2, 4, 6 and 12\,$\times$\,10$^{22}$\,cm$^{-2}$.}
    \label{fig:reg6_morph}
\end{figure*}

\begin{table*}
    \centering
    \begin{tabular}{l|cccccc}
    \hline
                    & Region\,1 & Region\,2 & Region\,3 & Region\,4 & Region\,5 & Region\,6 \\
    \hline \hline
      $^{12}$CO(3$-$2)     & 60$-$105 & 60$-$105 & 60$-$105 & 60$-$105 & 31$-$64  & 50$-$77 \\
      $^{13}$CO(3$-$2)     & 60$-$105 & 60$-$105 & 60$-$105 & 60$-$105 & 31$-$64 & 50$-$77 \\
      HCN(1$-$0)           & 60$-$105 & 60$-$105 & 60$-$105 & 60$-$107 & 25$-$69 & 30$-$80 \\
      HCO$^+$(1$-$0)       & 60$-$105 & 70$-$83 & 73.5$-$100.5 & 72$-$96.5 & 30.5$-$74 & 39.5$-$81 \\
      N$_2$H$^+$(1$-$0)    & 60$-$105 & 60$-$105 & 60$-$105 & 60$-$105 & 31$-$64 & 50$-$77 \\
      \hline
    \end{tabular}
    \caption{Velocity limits for the calculations of integrated intensity maps. Only the line components associated with the interarm area (for Regions\,1$-$4) and the Sagittarius arm (for Regions\,5$-$6) were considered here. The units are km\,s$^{-1}$ in each case.}
    \label{tab:integration}
\end{table*}

We created integrated intensity maps (henceforth $W$(Q) where Q is the molecular species and transition) of the studied molecular line emissions by taking the 0th moment of the spectra on each pixel. To define the velocity limits of the integration, we checked all spectral cubes to determine the exact velocities at which the tracer emission appears. These velocities were found in good agreement for the three IRAM-observed DGTs, and in case not, e.g. for the HCO$^+$(1$-$0) lines that are generally much narrower than the lines of the other species, the final integration limits were chosen to encompass all emission by the respective velocity component of the tracer while minimising the noise integrated in. The emission of $^{13}$CO(3$-$2) is generally observed in the same velocity ranges as the N$_2$H$^+$(1$-$0) line, and for the calculation of $W$($^{12}$CO(3$-$2)) the same limits were used as for $W$($^{13}$CO(3$-$2)), noting that these lines often show larger linewidths and blended velocity components. After this, the images were resampled to the same coordinate system and the same pixel size of 7$\farcs$61 (the original pixel size of the CHIMPS $^{13}$CO maps) so the detected emissions by different telescopes could be compared pixel by pixel. Table~\ref{tab:integration} shows the integration limits used for each line, and the integrated intensity maps are included in Figures \ref{fig:reg1_morph}$-$\ref{fig:reg6_morph}, alongside the $T_{\mathrm{dust}}$ maps from Hi-GAL. 

Region\,1 in the interarm has two bright local maxima close to each other on the $^{12}$CO map, neither of them coinciding with the $N$(H$_2$)$_{\mathrm{dust}}$ maxima. There is also a weaker, more nebulous emission area to the north. This morphology is followed by the $^{13}$CO and HCO$^+$ emission, but the northern nebulous area is not significant on the HCN and N$_2$H$^+$ maps. The various local $N$(H$_2$)$_{\mathrm{dust}}$ peaks appear with varied brightness in the emission of the different tracer species. Most of the CO-bright areas are relatively warm, with a CO-dark but N$_2$H$^+$-bright spot to the south-west that is the coldest. Region\,2 shows a bright clump to the north on all maps except N$_2$H$^+$ where it is weak. This clump roughly coincides with a peak in column density. There is a more nebulous, $^{13}$CO-bright area to the south which is even denser. The HCO$^+$ emission appears relatively bright on large portions of the observed map. Region\,3 shows higher overall $N$(H$_2$)$_{\mathrm{dust}}$ than the previous observing regions and has one large, bright clump in the emission of all species, which is mostly cold and also has high column density. Other smaller bright areas appear with varying brightness on the rest of the maps. The densest observing region of the interarm GMF, Region\,4, shows one bright clump in the emission of all studied species, which is dense and warm. A cold area to the north is bright in $^{13}$CO and appears in N$_2$H$^+$ as well, while no significant peak of HCO$^+$ and $^{12}$CO can be found there.

Region\,5 in the arm filament has the highest peak $N$(H$_2$)$_{\mathrm{dust}}$ both in the arm GMF and overall. Its structure is filamentary on the $N$(H$_2$)$_{\mathrm{dust}}$ map with denser clumps along the ridge that vary in temperature, but most of them are traced by all three of our IRAM-observed DGTs to some measure. The densest arc of cold material to the west is also traced by HCN and HCO$^+$, but not by N$_2$H$^+$. Region\,6, even denser overall than Region\,5, is an area where the arm GMF turns from west-east orientation to north-south. It appears as a large, wide arc of emission on the $^{12}$CO map. Two bright clumps, one to the east, which is warm, and one to the west, which is colder and more diffuse, are both traced by the DGTs. The warm clump is brighter in CO isotopologues, while the colder, more diffuse clump is brighter in the emission of HCN, HCO$^+$, and N$_2$H$^+$.

\subsection{Filling factors and cumulative fractions}
\label{sec:covandsum}

\begin{table*}
    \centering
    \begin{tabular}{l | c c c  c | c | c  c | c  }
    \hline
              & Region\,1 & Region\,2 & Region\,3 & Region\,4 & Interarm & Region\,5 & Region\,6 & Arm \\
        \hline
        \hline
         $^{12}$CO(3$-$2)   & 94.3 & 75.1 & 83.6 & 98.7 &    89.7 &    85.7 & 90.1 &   87.4 \\
         $^{13}$CO(3$-$2)   & 20.9 & 14.5 & 1.8 & 11.8 &     9.7  &    8.0 & 11.7  &   9.4 \\
         HCN(1$-$0)         & 12.6 & 14.4 & 4.1 & 21.2 &     12.7 &    11.8 & 13.4 &   12.4 \\
         HCO$^+$(1$-$0)     & 32.9 & 10.7 & 3.8 & 29.5 &     18.1 &    22.2 & 7.3  &   16.4 \\ 
         N$_2$H$^+$(1$-$0)  & 5.1  & 3.6  & 4.9 & 11.4 &     7.1  &    0.3 & 3.4   &   1.5  \\
         \hline
         A$_{\mathrm{tot}}$ [arcmin$^2$]  & 86.7 & 66.8 & 221.9 & 211.2 & 586.6 & 352.2 & 226.2 & 578.4 \\
         \hline \hline
        
    \end{tabular}
    \caption{Filling factors (\% of all pixels) of the observed species, and total sizes covered by the IRAM observations, for the regions separately and also summed up for the two filaments.}
    \label{tab:filling}
\end{table*}

We compute the global filling factors of the studied species as the ratio of pixels on which the integrated intensity of the line was detected with an emission higher than 3\,$\times$\,$\sigma_{\mathrm{int}}$ and the total number of pixels with measurements on the images, where $\sigma_{\mathrm{int}}$ is the standard deviation of values measured on an emission-free area on the respective integrated intensity images. Table~\ref{tab:filling} shows the computed filling factors as percentages for the six observing regions and summed up for the two GMFs. We note that this ``filling factor'' is different from the parameter calculated for the N$_2$H$^+$(1$-$0) line in \citet{feher2024} which was based on the pixels where the entire HFS of the line could be well fitted and its main group was detected with an integrated intensity of an S/N\,>\,3. In our two GMFs, the filling factor of N$_2$H$^+$(1$-$0) is lowest, and lower in the spiral arm than in the interarm with 1.5\% versus 7.1\%. The filling factor of HCO$^+$(1$-$0) is comparable to that of HCN(1$-$0), while the filling factor of $^{13}$CO(3$-$2) is generally lower than those. The filling factor of $^{12}$CO(3$-$2) is the highest, above 85\% in both GMFs.

\begin{figure*}
    \centering
    \includegraphics[width=\linewidth]{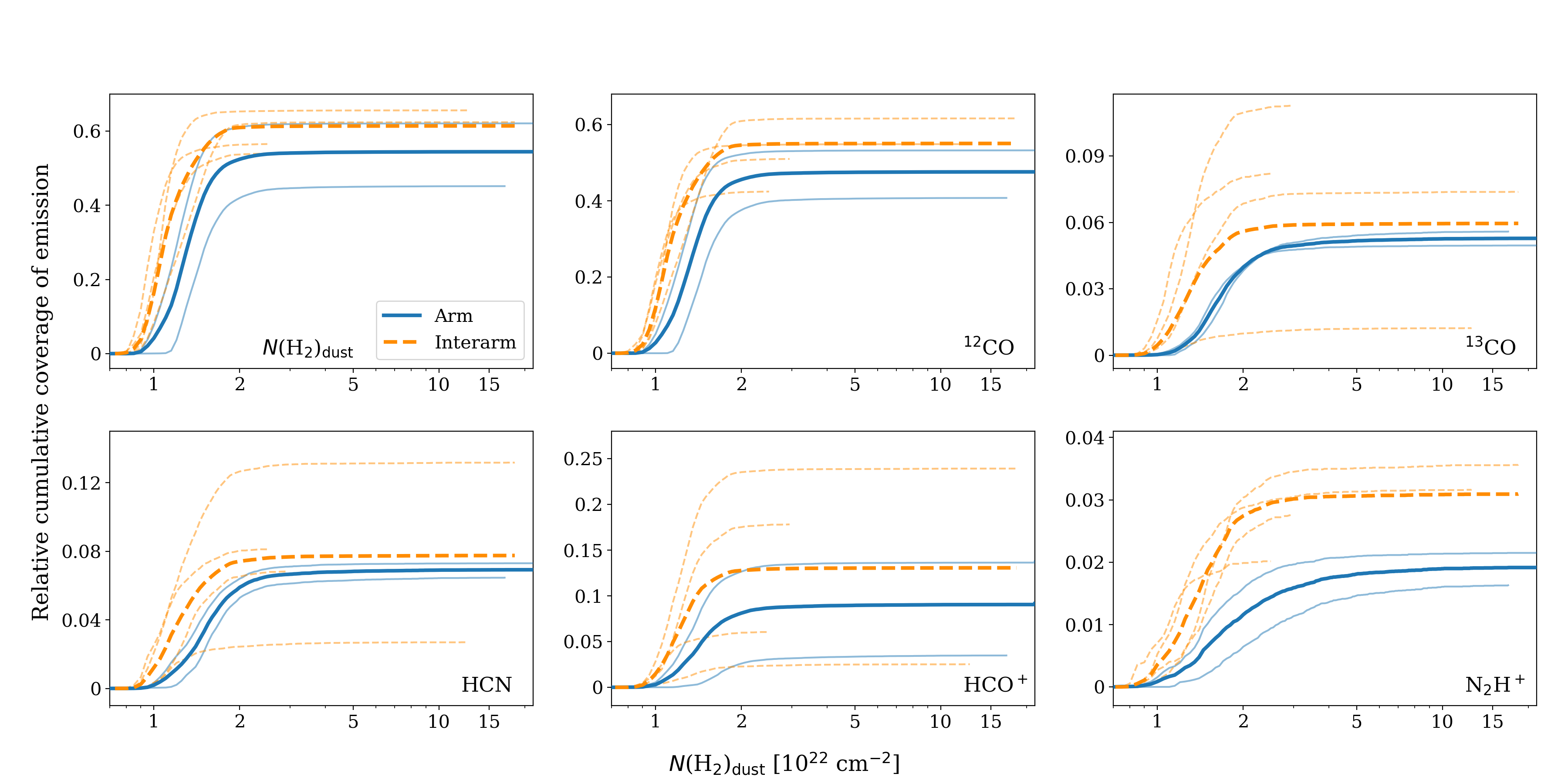}
    \caption{Relative cumulative coverage of the emission from the studied species in the two GMFs and the observed regions within. The thick blue and orange lines show the average curves for the arm and the interarm, respectively, computed by summing up the values in their respective observing regions. Fainter blue and dashed orange lines show the individual curves of the observing regions.}
    \label{fig:cumucov_species}
\end{figure*}

We can also examine the filling factor of the species as a function of column density. We define $N$(H$_2$)$_{\mathrm{dust}}$ bins from 0 to the $N$(H$_2$)$_{\mathrm{dust}}$ maxima in the respective observing regions with bin widths of 5\,$\times$\,10$^{20}$\,cm$^{-2}$. For each of the species and for each defined bin, we cumulatively count the number of pixels where the integrated intensity of the emission is higher than 3$\sigma_{\mathrm{int}}$. Fig.~\ref{fig:cumucov_species} shows the so-called relative cumulative coverage computed this way for each species and $N$(H$_2$)$_{\mathrm{dust}}$ bin in each observing region, relative to the full size of the respective observing regions. All the curves calculated for the arm GMF are lower than those of the interarm, and the difference between the two is most significant for N$_2$H$^+$. The N$_2$H$^+$ curve for the arm GMF reaches 50\% the maximum value at the highest column densities, showing a more gradual increase than the curves of the other species. Even if its overall filling factor is small, the N$_2$H$^+$ emission covers higher column density areas better than any of the other species. In contrast, the curves of $^{12}$CO rise steeply and reach their maximum at low $N$(H$_2$)$_{\mathrm{dust}}$, meaning $^{12}$CO is more likely to trace lower column density structures. For $^{13}$CO, HCN, and HCO$^+$, the curves obtained for the interarm and the arm are very close to each other, however, the individual curves of the various interarm observing regions show a large scatter. The coverage of these species in the different interarm regions is strongly variable and not very distinct from that of the arm regions.

\begin{figure*}
    \centering
    \includegraphics[width=\linewidth]{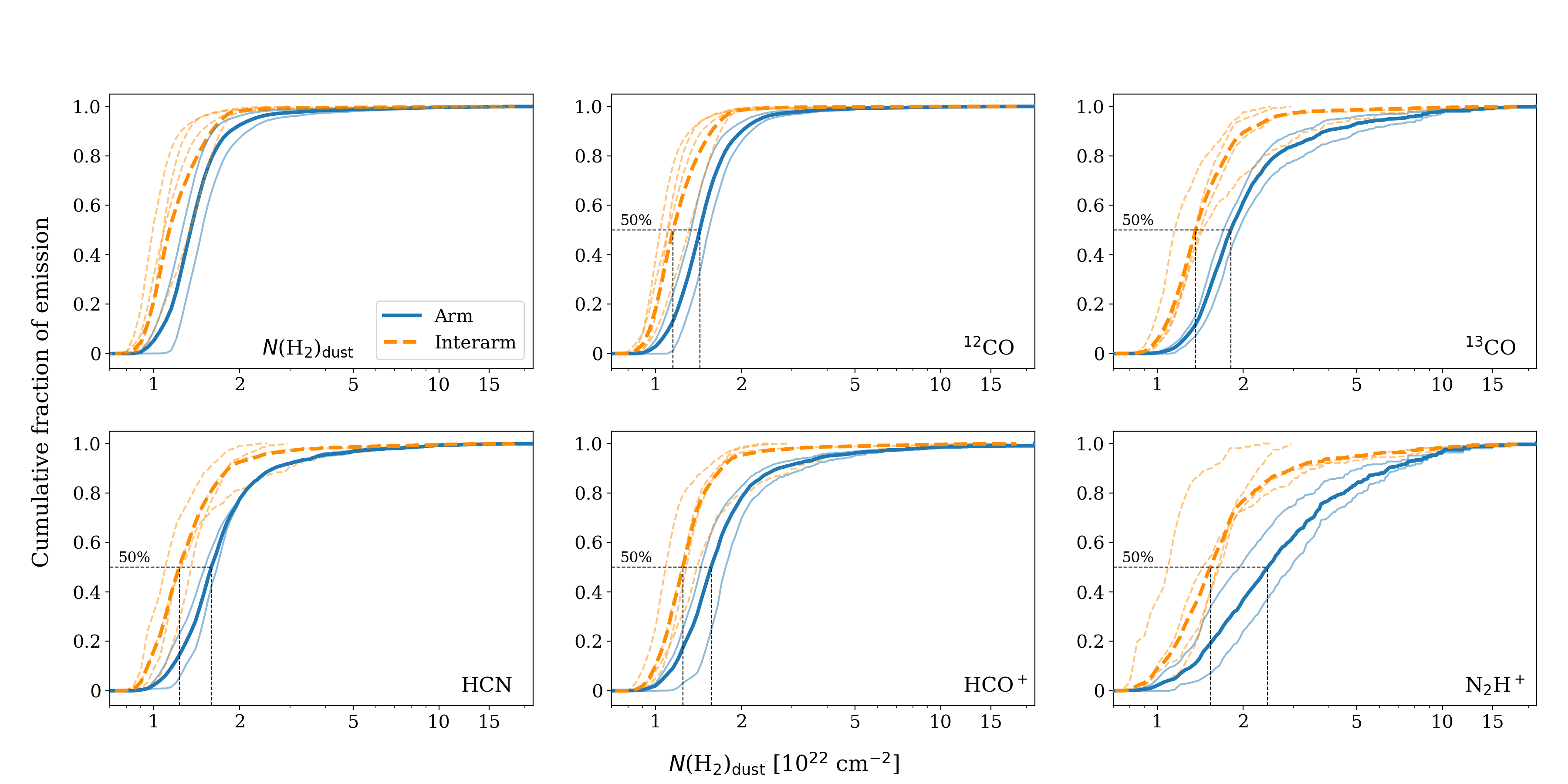}
    \caption{Cumulative fraction of the emission from the studied species in the two GMFs and the observing regions within. The legend of the image is the same as for Fig.~\ref{fig:cumucov_species}. Dashed black lines point out the $N$(H$_2$)$_{\mathrm{char}}$ values for both the arm and the interarm.}
    \label{fig:cumusum_species}
\end{figure*}

\begin{figure}
\centering
    \includegraphics[width=\linewidth]{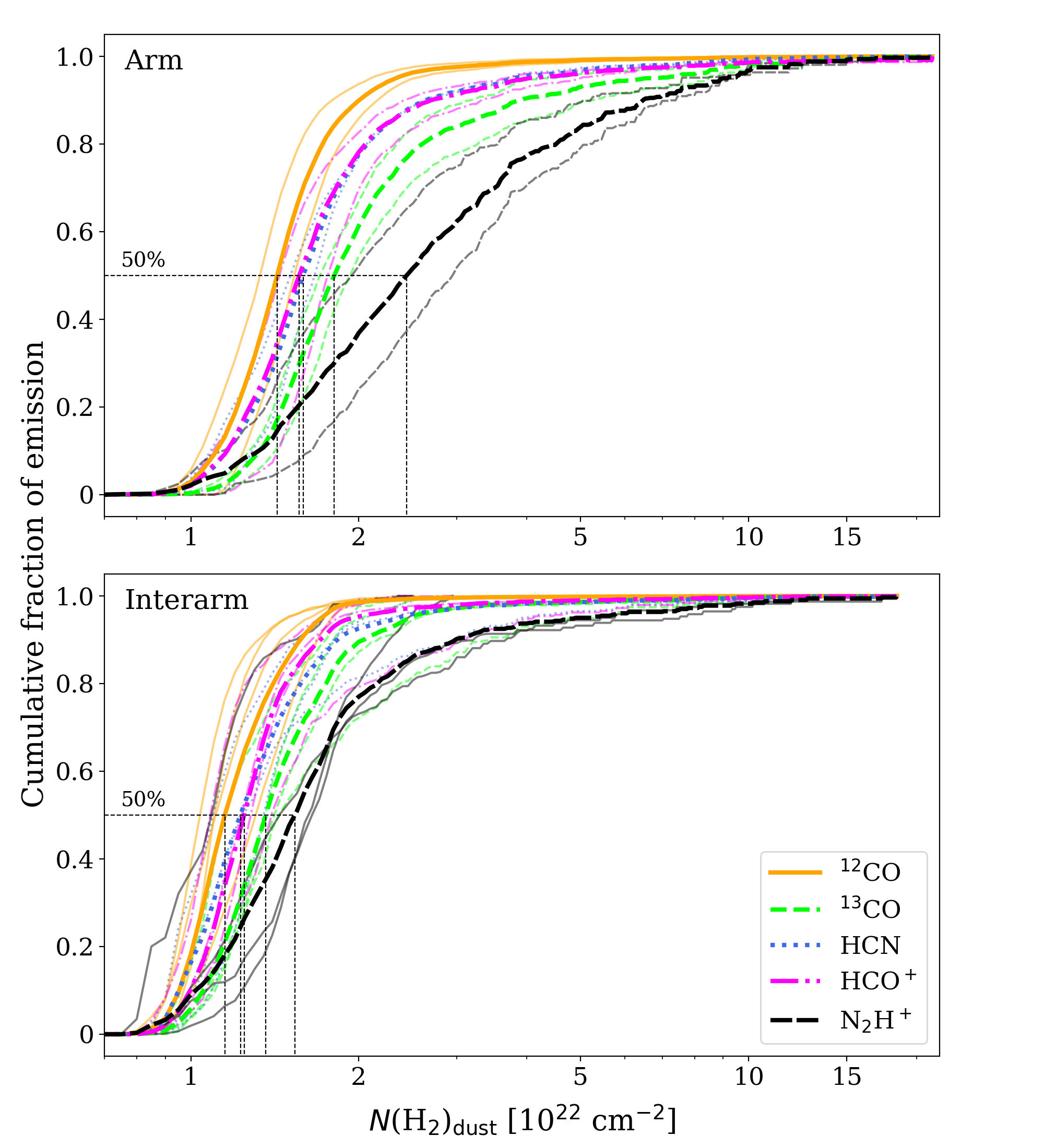}
    \caption{Cumulative fraction of the emission from the studied species in the two GMFs and the observing regions within. Thick lines with different colours and linestyles mark the average curves of the different species in the arm (top) and the interarm (bottom). Fainter lines mark the individual curves of the observing regions.}
    \label{fig:cumusum_species1}
\end{figure}

Fig.~\ref{fig:cumusum_species} shows the cumulative fraction of the emission from each studied species. To derive this, the same $N$(H$_2$)$_{\mathrm{dust}}$ bins and filtering below 3\,$\sigma_{\mathrm{int}}$ were used as for the calculation of the relative cumulative coverage. The curves of each species in the arm GMF are seen systematically shifted towards higher column densities, since the arm GMF is generally denser than the interarm GMF. Of all the observed species, N$_2$H$^+$, and especially N$_2$H$^+$ in the arm, is present in the least amounts at low column densities but traces the higher density areas better. We quantify this by following the example of \citet{kauffmann2017} and defining $N$(H$_2$)$_{\mathrm{char}}$ characteristic column density as a value below which 50\% of the line emission originates. $N$(H$_2$)$_{\mathrm{char}}$ is higher for the arm than for the interarm for all investigated species, with the greatest difference for N$_2$H$^+$, where it gives 1.5\,$\times$\,10$^{22}$\,cm$^{-2}$ in the interarm and 2.4\,$\times$\,10$^{22}$\,cm$^{-2}$ in the arm. Fig.~\ref{fig:cumusum_species1} allows us to compare the average cumulative fraction curves of the species in the arm and the interarm, respectively. In both filaments, $^{12}$CO shows the steepest curve reaching its maximum at the lowest column densities, while for N$_2$H$^+$, more of the emission originates from high density structures. The curves of HCN and HCO$^+$ are close to each other, and $^{13}$CO is intermediate between HCN, HCO$^+$ and N$_2$H$^+$. The difference between the $N$(H$_2$)$_{\mathrm{char}}$ of $^{12}$CO and N$_2$H$^+$ (the lowest and highest in both filaments) is smaller in the interarm (3\,$\times$\,10$^{21}$\,cm$^{-2}$) than in the arm (1.0\,$\times$\,10$^{22}$\,cm$^{-2}$), or in other words, the tracers show more distinct behaviour in the arm filament than in the interarm. Both the arm and interarm GMF curves are similar to those derived for Orion\,A \citep{kauffmann2017} and W49 \citep{barnes2020} in that the highest $N$(H$_2$)$_{\mathrm{char}}$ is shown by N$_2$H$^+$ and the lowest by $^{12}$CO, even though the studied transition of the CO isotopologues here is the J\,=\,3$-$2 and not the J\,=\,1$-$0 that are compared to N$_2$H$^+$(1$-$0) in the Orion\,A study. The differences between the lowest and highest $N$(H$_2$)$_{\mathrm{char}}$ in Orion\,A (10$^{22}$\,cm$^{-2}$) and W49 (around 8\,$\times$\,10$^{21}$\,cm$^{-2}$) make these areas more similar to our arm GMF than to the interarm in this regard.

\subsection{Molecular ratios}
\label{res:ratios}

\begin{table*}
    \centering
    \begin{tabular}{l| cccc| c | cc | c}
    \hline
         & Region\,1 & Region\,2 & Region\,3 & Region\,4 & Interarm & Region\,5 & Region\,6 &  Arm \\
         \hline
         \hline
    HCN/$^{13}$CO        & 0.3\,$\pm$\,0.1  & 0.5\,$\pm$\,0.2 & 0.3\,$\pm$\,0.1 & 0.3\,$\pm$\,0.1 &  0.3\,$\pm$\,0.2 & 0.4\,$\pm$\,0.2 & 0.4\,$\pm$\,0.2 & 0.4\,$\pm$\,0.2 \\
    HCO$^+$/$^{13}$CO    & 0.3\,$\pm$\,0.1  & 0.4\,$\pm$\,0.2 & 0.3\,$\pm$\,0.1 & 0.2\,$\pm$\,0.1 &  0.3\,$\pm$\,0.1 & 0.3\,$\pm$\,0.1 & 0.4\,$\pm$\,0.2 & 0.3\,$\pm$\,0.2\\
    N$_2$H$^+$/$^{13}$CO & 0.2\,$\pm$\,0.1  & 0.5\,$\pm$\,0.2 & 0.4\,$\pm$\,0.3 & 0.4\,$\pm$\,0.3 &  0.4\,$\pm$\,0.2 & 0.2\,$\pm$\,0.2 & 0.2\,$\pm$\,0.1 & 0.2\,$\pm$\,0.1\\
    HCO$^+$/HCN          & 1.1\,$\pm$\,0.4  & 0.9\,$\pm$\,0.3 & 0.8\,$\pm$\,0.2 & 0.8\,$\pm$\,0.4 &  0.9\,$\pm$\,0.4 & 0.8\,$\pm$\,0.4 & 1.2\,$\pm$\,0.5 & 0.9\,$\pm$\,0.5\\
    N$_2$H$^+$/HCN       & 0.9\,$\pm$\,0.4  & 1.3\,$\pm$\,0.6 & 1.1\,$\pm$\,0.6 & 1.2\,$\pm$\,0.6 &  1.1\,$\pm$\,0.6 & 0.5\,$\pm$\,0.2 & 0.7\,$\pm$\,0.4 & 0.6\,$\pm$\,0.3\\
    N$_2$H$^+$/$N$(H$_2$)$_{\mathrm{dust}}$  & 0.01\,$\pm$\,0.004 & 0.03\,$\pm$\,0.01 & 0.02\,$\pm$\,0.02 & 0.3\,$\pm$\,0.01 & 0.02\,$\pm$\,0.01 & 0.01\,$\pm$\,0.005 & 0.01\,$\pm$\,0.01 & 0.01\,$\pm$\,0.01\\
        \hline
    \end{tabular}
    \caption{Average molecular ratios detected in the two GMFs and the observing regions within. The unit of N$_2$H$^+$/$N$(H$_2$)$_{\mathrm{dust}}$ is K\,km\,s$^{-1}$/10$^{26}$cm$^{-2}$.}
    \label{tab:allregion_ratios}
\end{table*}

\begin{figure*}
    \centering
    \includegraphics[width=\linewidth]{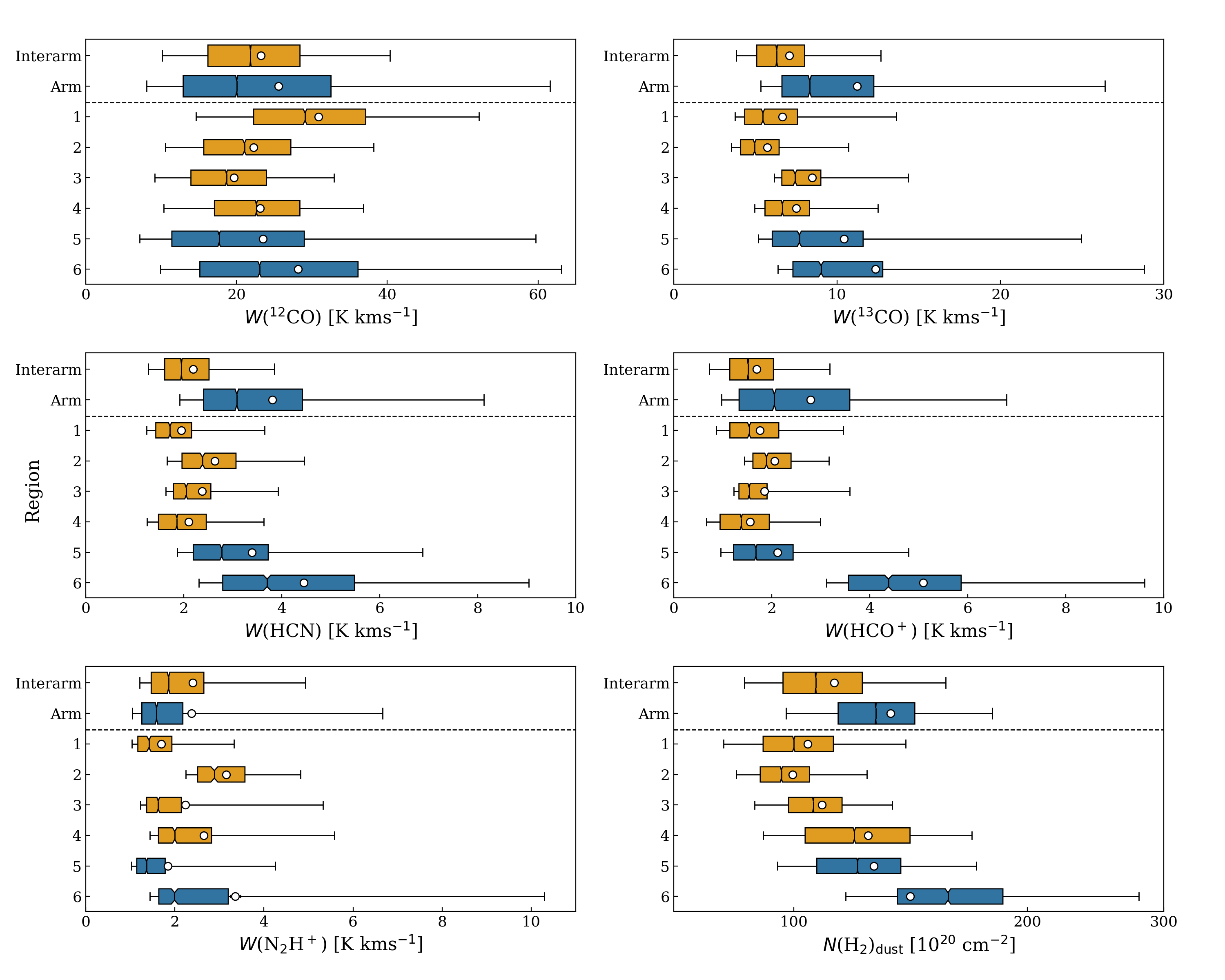}
    \caption{The detected molecular emission levels in the two GMFs and the observing regions within. Each box extends from the first quartile to the third quartile with a vertical line at the median value and a notch representing the confidence interval of the median. The whiskers extend from the 5\% percentile to the 95\% percentile of the data. Blue colour corresponds to the arm and orange to interarm. White circles mark average values.}
    \label{fig:allregion_emissions_box1}
\end{figure*}

\begin{figure*}
    \centering
   \includegraphics[width=\linewidth]{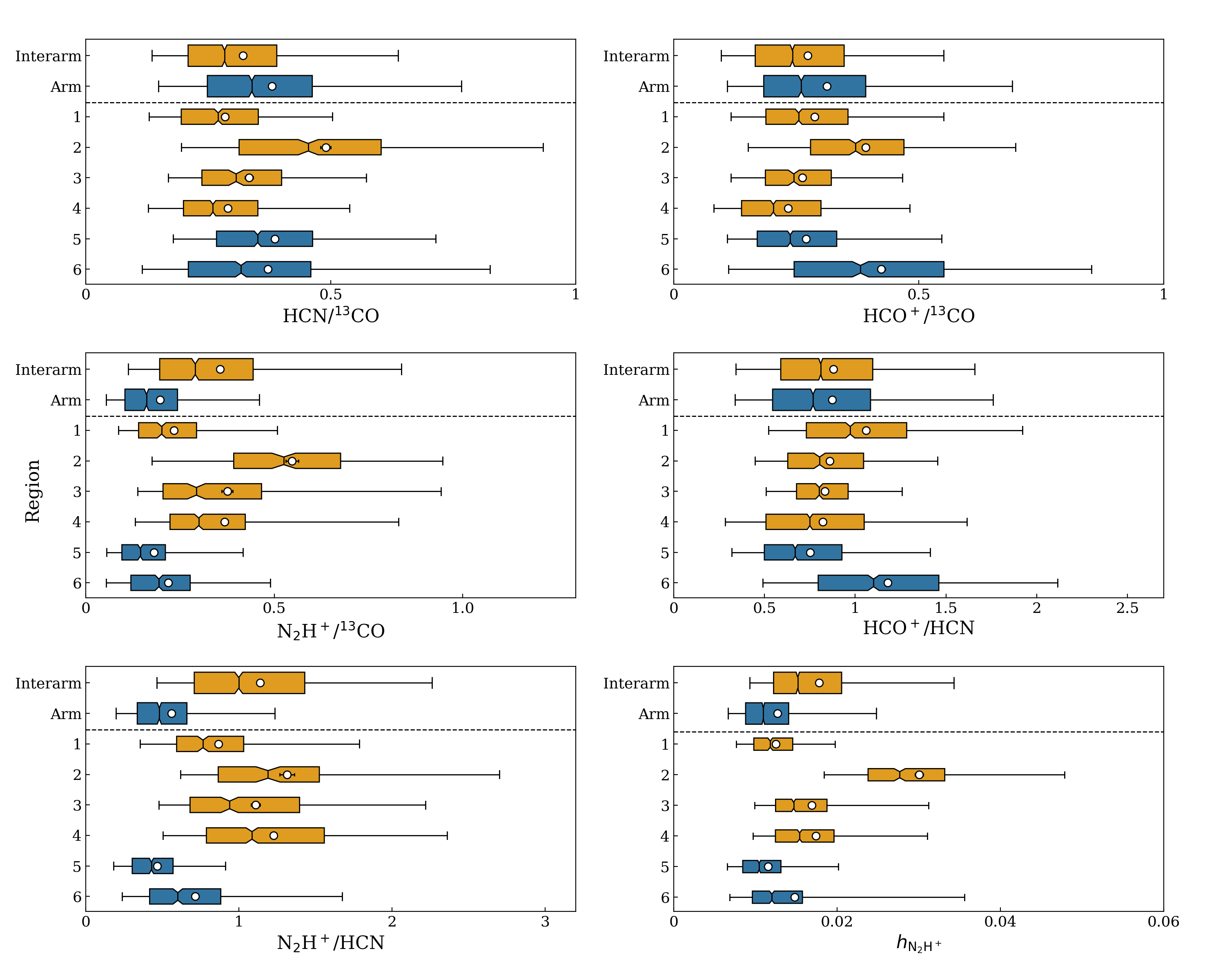}
    \caption{The detected molecular ratios in the two GMFs and the observing regions within. The boxes are similarly plotted as on Fig.~\ref{fig:allregion_emissions_box1}.}
    \label{fig:allregion_emissions_box2}
\end{figure*}

We calculate the ratios HCN/$^{13}$CO, HCO$^+$/$^{13}$CO, N$_2$H$^+$/$^{13}$CO \citep[the parameter called $f'_{\mathrm{DG}}$ in][]{feher2024}, HCO$^+$/HCN, N$_2$H$^+$/HCN, and N$_2$H$^+$/$N$(H$_2$)$_{\mathrm{dust}}$ \citep[the parameter called $h_{\mathrm{N_2H^+}}$ in][]{feher2024} using the integrated intensity values of the observed species in each observing region on pixels where the integrated intensity of the relevant species were detected with at least 2\,$\times$\,$\sigma_{\mathrm{int}}$. Figures \ref{fig:reg1_ratio}$-$\ref{fig:reg6_ratio} show the molecular ratio maps and Table~\ref{tab:allregion_ratios} collects the average values of the molecular ratios in the observing regions and the two GMFs. In Fig.~\ref{fig:allregion_emissions_box1}-\ref{fig:allregion_emissions_box2}, the observed ranges of molecular emissions and molecular ratios are illustrated with box plots.

We already saw in \citet{feher2024} that the average $N$(H$_2$)$_{\mathrm{dust}}$ level is higher and the $^{13}$CO emission is brighter in the arm than in the interarm. A similar difference can be observed in the filament averages of the HCN and HCO$^+$ emission. In contrast, there is little difference in the average N$_2$H$^+$ emission between the two GMFs, and only a small difference in the $^{12}$CO averages. The observed difference in the emission levels of HCO$^+$, N$_2$H$^+$ and the H$_2$ column densities between the various arm and interarm observing regions is mostly on account of Region\,6, the southern, north-south oriented half of the arm GMF. This region is the brightest in all observed molecular emissions and the densest on the Hi-GAL maps. For N$_2$H$^+$ and to some extent even for $^{13}$CO and HCN, the bulk of the molecular emission (the interquartile range in the box plots of Fig.~\ref{fig:allregion_emissions_box1}) in Region\,6 is in a similar interval to the interarm regions, with localized, very bright peaks causing a shift in the average. For HCO$^+$ and $N$(H$_2$)$_{\mathrm{dust}}$, already the interquartile range of Region\,6 is higher than for the interarm regions. The other arm filament section, Region\,5, is very similar in molecular emission and H$_2$ column density to the interarm, except localized peaks in $^{13}$CO and HCN emission. Regarding the $^{12}$CO, HCO$^+$, and N$_2$H$^+$ emission, there are often larger variations in the emission levels between regions of the same GMF than between the two GMFs.

We investigated the N$_2$H$^+$/$^{13}$CO ratio in \citet{feher2024} and saw that the arm GMF shows lower values of this ratio with an average of 0.2\,$\pm$\,0.1 in contrast with the interarm average of 0.4\,$\pm$\,0.2. As seen above, this mostly reflects the variation of the $^{13}$CO emission level between the two GMFs, not the N$_2$H$^+$. Examining Fig.~\ref{fig:allregion_emissions_box2}, we see a similar trend in the global average N$_2$H$^+$/HCN ratio with the interarm average being nearly twice the arm average. The variation of this ratio is governed by the brighter HCN emission in the arm, while the N$_2$H$^+$ emission varies little. On the other hand, the HCO$^+$/HCN ratio is remarkably constant (both the filament average and the interquartile range) across the two GMFs, while variations among the different observing regions can be seen. Regarding the HCN/$^{13}$CO and HCO$^+$/$^{13}$CO ratios, slightly higher values appear in the arm filament, mostly on account of the increase in the $^{13}$CO level. As we have seen in \citet{feher2024}, the N$_2$H$^+$/$N$(H$_2$)$_{\rm dust}$ ratio ($h_{\mathrm{N_2H^+}}$) shows a similar variation to the N$_2$H$^+$/$^{13}$CO ratio.

\begin{figure}
    \centering
    \includegraphics[width=\linewidth]{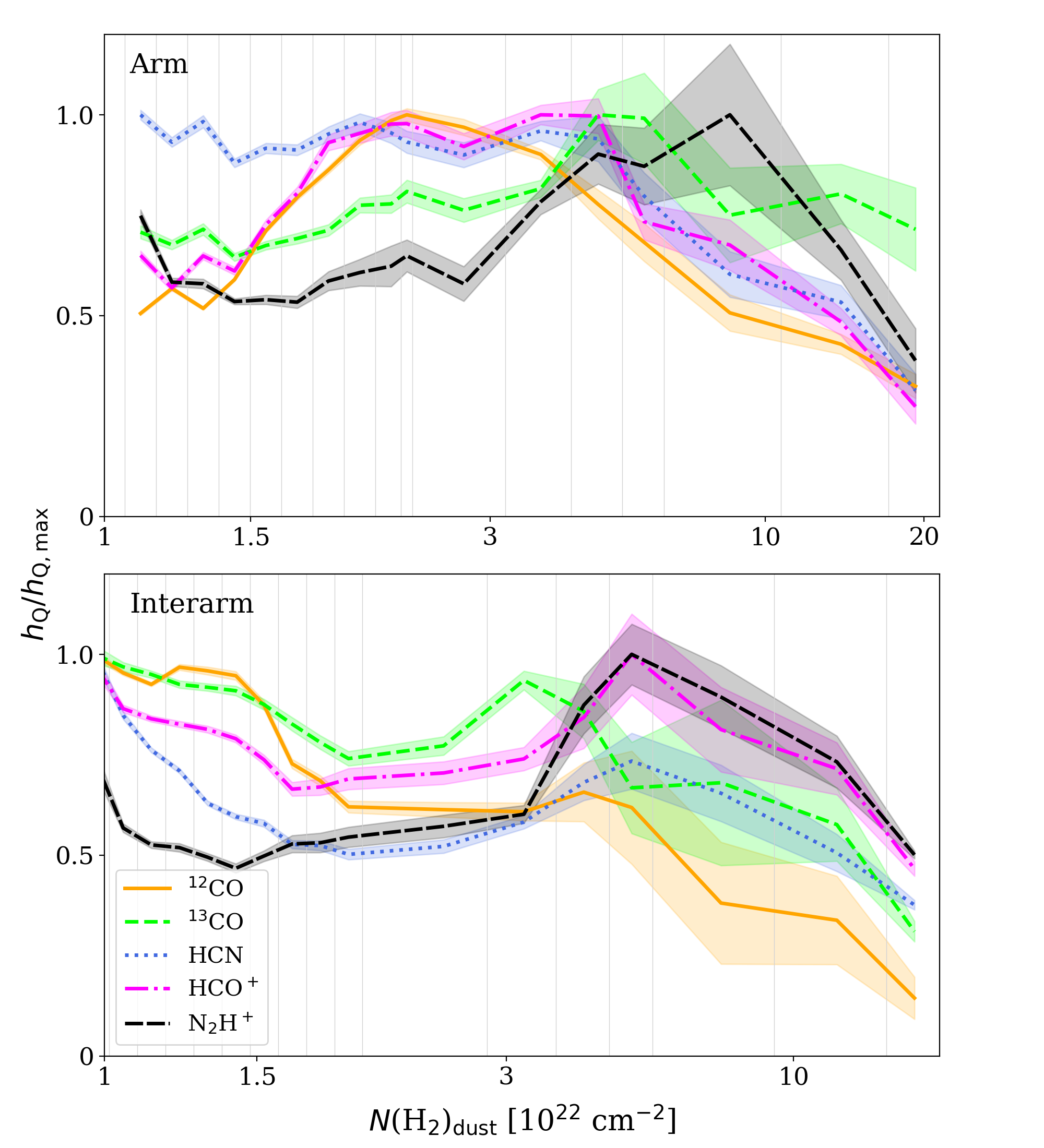}
    \caption{The line-to-mass ratio of the studied molecular species in the two GMFs expressed as $h_{\mathrm{Q}}$/$h_{\mathrm{Q, max}}$ as a function of H$_2$ column density.}
    \label{fig:hq}
\end{figure}

We examine this last parameter, the so-called line-to-mass ratio, for the other species as well. We define H$_2$ column density bins with limits at percentages of the $N$(H$_2$)$_{\mathrm{dust}}$ maxima inside the two GMFs, where the percentage ranges are: 5$-$10\% by 0.5\%, 10$-$30\% by 5\%, 50, 80, and 100\%. This definition balances between sampling the column densities in the two GMFs well (since the column density ranges in the different observing regions are very different) while also having enough pixels in each bin for statistics. On each pixel in each bin defined like this, we calculate $h_{\mathrm{Q}}$\,=\,$W$(Q)/$N$(H$_2$)$_{\mathrm{dust}}$. A variation of this parameter using $A_{\mathrm{V}}$ extinction instead of $N$(H$_2$)$_{\mathrm{dust}}$ was used by e.g. \citet{kauffmann2017, barnes2020, patra2022} as a marker for how the emission from transition and species $Q$ relates to the mass reservoir characterised by $A_{\mathrm{V}}$. In those studies, most molecules (CO isotopologues, HCN, CN, HCO$^+$) start with a significant value of $h_{\mathrm{Q}}$ at low $A_{\mathrm{V}}$ and increase towards their maximum values around $A_{\mathrm{V}}$\,=\,5$-$20\,mag, then decrease again with further increasing $A_{\mathrm{V}}$. The molecules HCN and HCO$^+$ were found not as strongly decreasing with $A_{\mathrm{V}}$ as CO, and a single molecule, N$_2$H$^+$, showed a higher than zero $h_{\mathrm{Q}}$ only at $A_{\mathrm{V}}$\,>\,10 mag and steadily rising after. Fig.~\ref{fig:hq} shows the average line-to-mass ratio $h_{\mathrm{Q}}$ for the two GMFs, scaled with $h_{\mathrm{Q,max}}$ to values between 0 and 1 for easier comparison between species. 

In the arm, HCN has the highest line-to-mass ratio at low column densities which then decreases starting at 5\,$\times$\,10$^{22}$\,cm$^{-2}$. Both $h_{\mathrm{^{12}CO}}$ and the $h_{\mathrm{HCO^+}}$ trace the 1.7$-$5\,$\times$\,10$^{22}$\,cm$^{-2}$ column density range best, with $h_{\mathrm{^{12}CO}}$ falling off at somewhat lower values. The $h_{\mathrm{^{13}CO}}$ curve peaks around 5\,$\times$\,10$^{22}$\,cm$^{-2}$ but shows consistently high values throughout the density range. In a singular way, $h_{\mathrm{N_2H^+}}$ is very low up to 5\,$\times$\,10$^{22}$\,cm$^{-2}$ column densities, peaks at around 10$^{23}$\,cm$^{-2}$ (the latest), and decreases sharply after. In the interarm GMF, $^{12}$CO, $^{13}$CO, and HCN all start with high $h_{\mathrm{Q}}$ values, which then decrease towards higher densities, although $h_{\mathrm{^{13}CO}}$ and $h_{\mathrm{HCN}}$ also show secondary peaks around 3 and 5\,$\times$\,10$^{22}$\,cm$^{-2}$, respectively. Similarly to the arm GMF, $h_{\mathrm{N_2H^+}}$ rises late and peaks at the highest column densities (but somewhat lower than in the arm), around 5\,$\times$\,10$^{22}$\,cm$^{-2}$. In contrast with the arm, however, the $h_{\mathrm{HCO^+}}$ curve peaks alongside $h_{\mathrm{N_2H^+}}$, thus, the behaviour of the N$_2$H$^+$ emission is not as distinct. These results are largely consistent with the findings of other inner Galaxy studies of these species \citep[e.g.][]{kauffmann2017,barnes2020}. 

\subsection{Molecular emission inside dense clumps}

\begin{figure*}
    \centering
  \includegraphics[width=\linewidth]{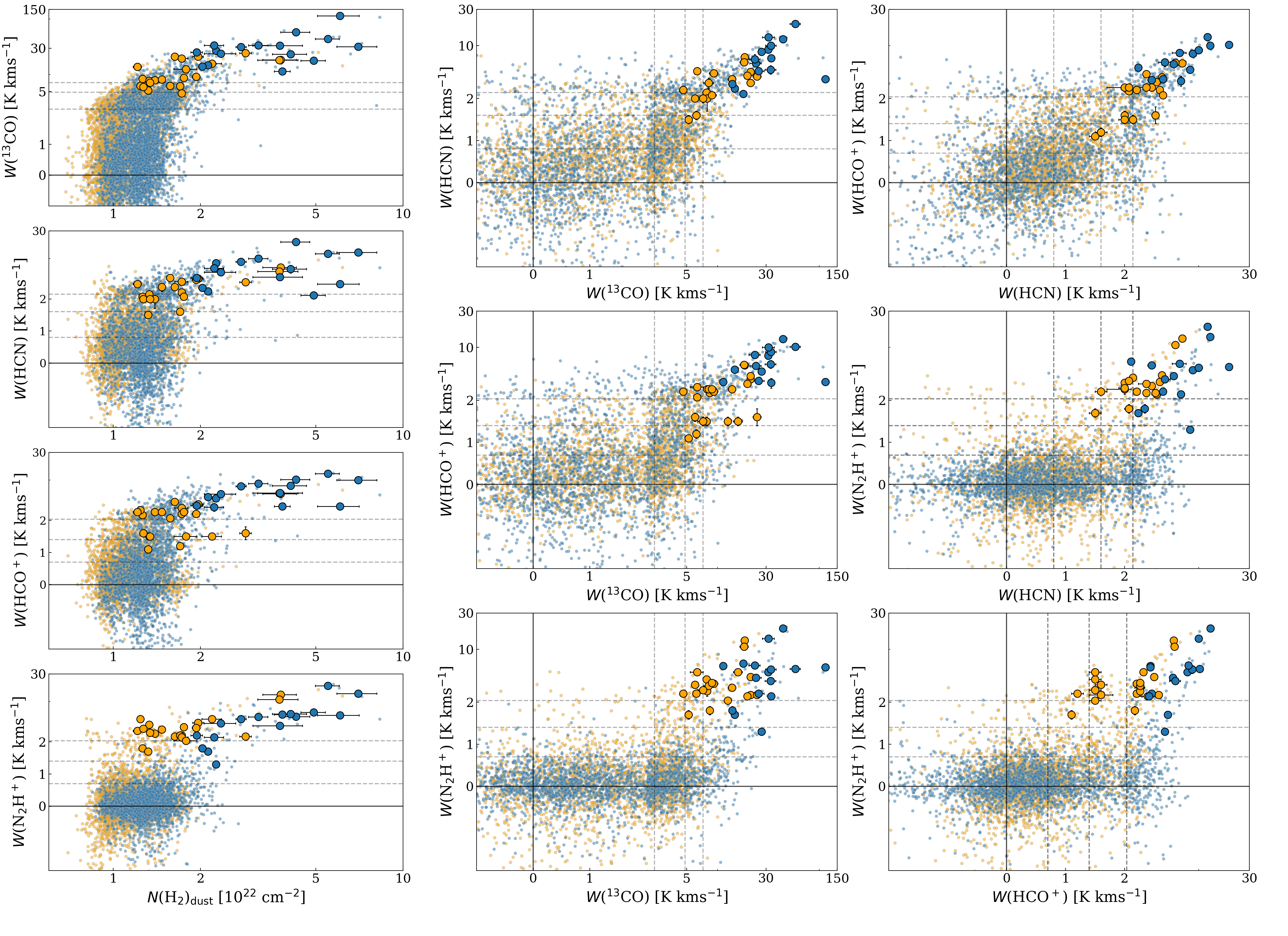}
    \caption{Correlations between the integrated intensities of the studied species in the two GMFs. Small circles mark the beam-averaged values and large circles mark the average values for the N$_2$H$^+$-clumps. Orange colour marks values measured in the interarm regions, and blue colour marks values measured in the arm regions. Left column: $N$(H$_2$)$_{\mathrm{dust}}$ versus $^{13}$CO(3$-2$), HCN, HCO$^+$, and N$_2$H$^+$ integrated intensity; Middle and right columns: the integrated intensities of every DGT as a function of each other. Solid black line marks the zero levels of molecular emission, and dashed lines mark average 1, 2, and 3\,$\times$\,$\sigma_{\mathrm{int}}$ measured on the respective integrated intensity maps.}
    \label{fig:gen_corr}
\end{figure*}

In \citet{feher2024} we defined so-called clusters or clumps using dendrogram clustering on the N$_2$H$^+$(1$-$0) spectral cubes. These clumps are parsec- and sub-parsec-size, largely elliptical regions with bright N$_2$H$^+$(1$-$0) emission and no significant sub-structure at this spatial resolution. We report the average integrated intensities of the emission from our studied molecular species (using only pixels above the 2\,$\times$\,$\sigma_{\mathrm{int}}$ level) and the average $N$(H$_2$)$_{\mathrm{dust}}$ and $T_{\mathrm{dust}}$ computed inside each of our N$_2$H$^+$-clumps in Table~\ref{tab:clusters}. We also include average values calculated over the six observed regions for comparison. Fig.~\ref{fig:gen_corr} shows the correlations between the integrated intensities of the different species (and $N$(H$_2$)$_{\mathrm{dust}}$) for every beam (using the IRAM\,30m beamwidth) in the two GMFs, and also the average values for the N$_2$H$^+$-clumps. From now on, we only study those N$_2$H$^+$-clumps which have been found at a $v_{\mathrm{LSR}}$ corresponding to that of their host filament, thus, 38 clumps from \citet{feher2024}, all except R1CL7 and R4CL8.

\begin{table*}
\footnotesize
    \centering
    \begin{tabular}{lllll}
    \hline
              & \multicolumn{1}{c}{HCN} & \multicolumn{1}{c}{HCO$^+$} & \multicolumn{1}{c}{N$_2$H$^+$} & \multicolumn{1}{c}{$N$(H$_2$)$_{\mathrm{dust}}$} \\
    \hline
    \hline 
    \makecell[cl]{$^{13}$CO} & \makecell[cl]{0.76 (2.5\,$\times$\,10$^{-8}$) \\ 0.50 (0.02) \\ 0.68 (4.0\,$\times$\,10$^{-3}$)}  &  \makecell[cl]{0.72 (4.5\,$\times$\,10$^{-7}$) \\ 0.53 (0.01) \\ 0.47 (0.07)}  & \makecell[cl]{0.37 (0.02) \\ 0.21 (0.35) \\ 0.21 (0.43)} & \makecell[cl]{0.77 (1.3\,$\times$\,10$^{-8}$) \\ 0.59 (4.0\,$\times$\,10$^{-3}$) \\ 0.42 (0.1)}  \\
    \hline
       \makecell[cl]{HCN}      & \makecell[cc]{$\times$} & \makecell[cl]{0.77 (1.8\,$\times$\,10$^{-8}$) \\ 0.61 (2.0\,$\times$\,10$^{-3}$) \\ 0.80 (1.9\,$\times$\,10$^{-4}$)} & \makecell[cl]{0.31 (0.06) \\ 0.16 (0.49) \\ 0.22 (0.41)} & \makecell[cl]{0.59 (8.6\,$\times$\,10$^{-5}$) \\ 0.45 (0.04) \\ 0.29 (0.27)} \\
       \hline
       \makecell[cl]{HCO$^+$}      & \makecell[cc]{$\times$} & \makecell[cc]{$\times$} & \makecell[cl]{0.49 (1.7\,$\times$\,10$^{-3}$) \\ 0.41 (0.06) \\ 0.41 (0.12)} & \makecell[cl]{0.58 (1.2$\times$\,10$^{-4}$) \\ 0.26 (0.24) \\ 0.42 (0.11)} \\
       \hline
       \makecell[cl]{N$_2$H$^+$}   & \makecell[cc]{$\times$} & \makecell[cc]{$\times$} & \makecell[cc]{$\times$} & \makecell[cl]{0.61 (5.6\,$\times$\,10$^{-5}$) \\ 0.30 (0.17) \\ 0.88 (6.1\,$\times$\,10$^{-6}$)} \\ 
       \hline
    \end{tabular}
    \caption{The computed Spearman rank-order correlation coefficients and p-values between the average molecular emissions and average column densities of N$_2$H$^+$-clumps in the two GMFs. The first row in a cell is between the two relevant values of all clumps, the second is only for interarm clumps, the third is only for arm clumps.}
    \label{tab:spearman}
\end{table*}

Generally, clumps in the arm GMF tend to have higher column densities, are brighter in $^{13}$CO emission, and are somewhat brighter in HCN emission than clumps in the interarm GMF, while there is little difference between the arm and interarm clumps regarding HCO$^+$ and N$_2$H$^+$ emission. To characterize the relations between the molecular emission of the studied species in the clumps, we computed Spearman rank-order correlation coefficients for each combination (Table~\ref{tab:spearman}). According to the results, there are fairly tight correlations (p-value\,>\,0.5) between the average H$_2$ column densities and the average molecular emission of clumps in general (i.e. arm and interarm clumps together) with the strongest relation being between $^{13}$CO brightness and $N$(H$_2$)$_{\mathrm{dust}}$. When considering the two GMFs separately, there is an especially strong correlation between $W$(N$_2$H$^+$) and $N$(H$_2$)$_{\mathrm{dust}}$ in the arm clumps, while interestingly, the relation between the same two quantities is quite weak for the interarm clumps. The correlation of HCN and HCO$^+$ average emissions is very strong in the clumps in general, and the $^{13}$CO emission is also well-correlated with the HCN and HCO$^+$ emission. The worst correlation is found between the average $W$($^{13}$CO) and $W$(N$_2$H$^+$), and between the average $W$(HCN) and $W$(N$_2$H$^+$) of the clumps, both in general as well as considering arm and interarm objects separately.

The studies by \citet{tafalla2021} and \citet{tafalla2023} investigate molecular line emission from Galactic clouds like California, Perseus, and Orion\,A. In their analysis using stratified random sampling, they also found strong correlations between molecular line intensities and H$_2$ column densities. In those studies, the intensity of CO isotopologues increases slower than linearly with $N$(H$_2$)$_{\mathrm{dust}}$ while most DGT intensities increase linearly. The largest deviation appears for N$_2$H$^+$ which shows a rapid transition from undetected to relatively bright and continues to increase close to linearly near 10$^{22}$\,cm$^{-2}$. The Orion\,A data also shows the molecular line emission dependence on gas temperature. For our data, we find the same type of tight correlations between column density and line emission for the clump averages, however, there is no significant difference in the linearity of the increase of line emissions with column density. We note that our ``low column densities'' are already above 10$^{22}$\,cm$^{-2}$. \citet{tafalla2021} also reproduces the type of plot we presented in Fig.~\ref{fig:hq}. Their results are very similar to the tendencies derived by \citet{kauffmann2017} and \citet{barnes2020} that we have discussed earlier.

\section{Discussion}
\label{sec:discuss}

\subsection{The dense gas through density regimes and environments in the Galaxy}

We have presented analysis of the molecular emission of DGTs from filamentary regions on parsec to tens of parsec scales. By examining contiguous structures over this range of scales, we bridge to similar spatial scales reached by high-resolution interferometric studies of nearby galaxies. In doing so, we can now explore whether the trends that are observed across entire galaxies emerge from self-similar trends on smaller scales or if there is significant variation within a ``beam'' that is averaged out within an extragalactic resolution element. While further resolving star forming regions in external galaxies to reveal smaller-than-beamsize effects governing the large-scale view of these clouds requires instrumental innovation, we can instead use GMFs in our own galaxy to shed light onto some of these spatial scale-dependent variations. Averaging the molecular emission for the six observing regions towards the two GMFs, one in the Sagittarius spiral arm and one in an interarm area, is analogous to resolving the GMF structure further. Figures \ref{fig:allregion_emissions_box1} and \ref{fig:allregion_emissions_box2} show us that there is often greater variation in the line intensities and ratios between regions \textit{within} an environment than there is \textit{between} the two environments. This suggests that, at least within the inner disc of the Milky Way, in most cases molecular emission and line ratio trends do not strongly depend on the ``environment'' as we have defined it here. 

In the following, we discuss these variations from the largest scales of our survey, best traced by CO, down to denser gas probed by the 3\,mm line observations and the variation in parameters as function of the environments (arm and interarm).

\subsubsection{Molecular gas tracers across the Galaxy in the literature}
\label{disc:gasingalaxy}

Based on the J\,=\,3$-$2 transitions of $^{12}$CO and $^{13}$CO, tracing the largest scales and the lowest column densities the best, and N$_2$H$^+$(1$-$0) which is reportedly most fitting to catch the densest structures, we saw variations of different magnitudes in our data. To put this in a galactic context, we turn to large surveys of CO isotopologues and sub-millimeter emission in the literature. \citet{romanduval2010} found the surface density of molecular clouds measured with the emission of $^{13}$CO(1$-$0) to steeply decline with Galactocentric radius in the GRS (Galactic Ring Survey), and the study of \citet{romanduval2016} shows that the dense gas mass fraction measured with $^{13}$CO versus $^{12}$CO\,+\,$^{13}$CO also declines with Galactocentric radius on kiloparsec scales. However, the dense gas mass fraction derived using the mass of sub-millimeter clumps shows no such variation \citep{battisti2014}. Recently published results by \citet{rigby2024} comparing survey data of the J\,=\,3$-$2 transitions of CO (the same transitions as in our data, which should trace higher critical density gas than the lower transitions) found that the CO-to-H$_2$ ratio, the so-called X-factor, does not vary between the inner and outer Galaxy, and the clump properties are also consistent at all Galactic radii. It has thus been suggested that once dense structures are formed in molecular clouds, their characteristics may be governed by smaller-scale, internal properties (nonetheless perhaps inherited from their environments) rather than large-scale tendencies.

Regarding the relation between the dense gas content and star formation of spiral arms and interarm areas, studies have indicated that although material is concentrated in the arms, there is no compelling evidence that SFE varies in association with them \citep{moore2012, eden2013}. The conclusion was also supported by the SFF (star-forming fraction, a parameter correlated with SFE) tendencies determined by \citet{ragan2016} only showing the spiral arms as features in the overall distribution of their sources but not in the derived values. This again suggests that on the scales of parsecs to tens of parsecs, the gas traced by these transitions is not strongly affected by the galaxy-scale environmental conditions that govern or derive from spiral arm formation. In recent DGT pointing surveys of nearby galaxies, HCN emission shows a good correlation with N$_2$H$^+$ emission on kiloparsec scales \citep{jimenez2023}, but by resolving molecular clouds further, environmental differences can be observed \citep{stuber2023}, possibly indicating changes in SFE -- one of the points that motivated our study.

\subsubsection{Molecular gas tracers in our data on filament- and clump-scales and across two environments}

Our data does not allow for conclusions as a function of Galactic radius, nevertheless, it is clear that the variation seen in our measures of dense gas mass fraction will strongly depend on the tracers we use and the scales we resolve. In the following, we will use statistical analyses on the observed emission levels and line ratios between the arm and interarm environments to quantify similarities and differences. We will consider this on both the whole filament scale and on the clump scale (where the clumps are our defined N$_2$H$^+$-clumps) in order to assess whether the sufficiently dense gas assumed to be tightly connected to star formation behaves differently from the bulk of the gas in these structures.

\begin{figure*}
    \centering
    \includegraphics[width=\linewidth]{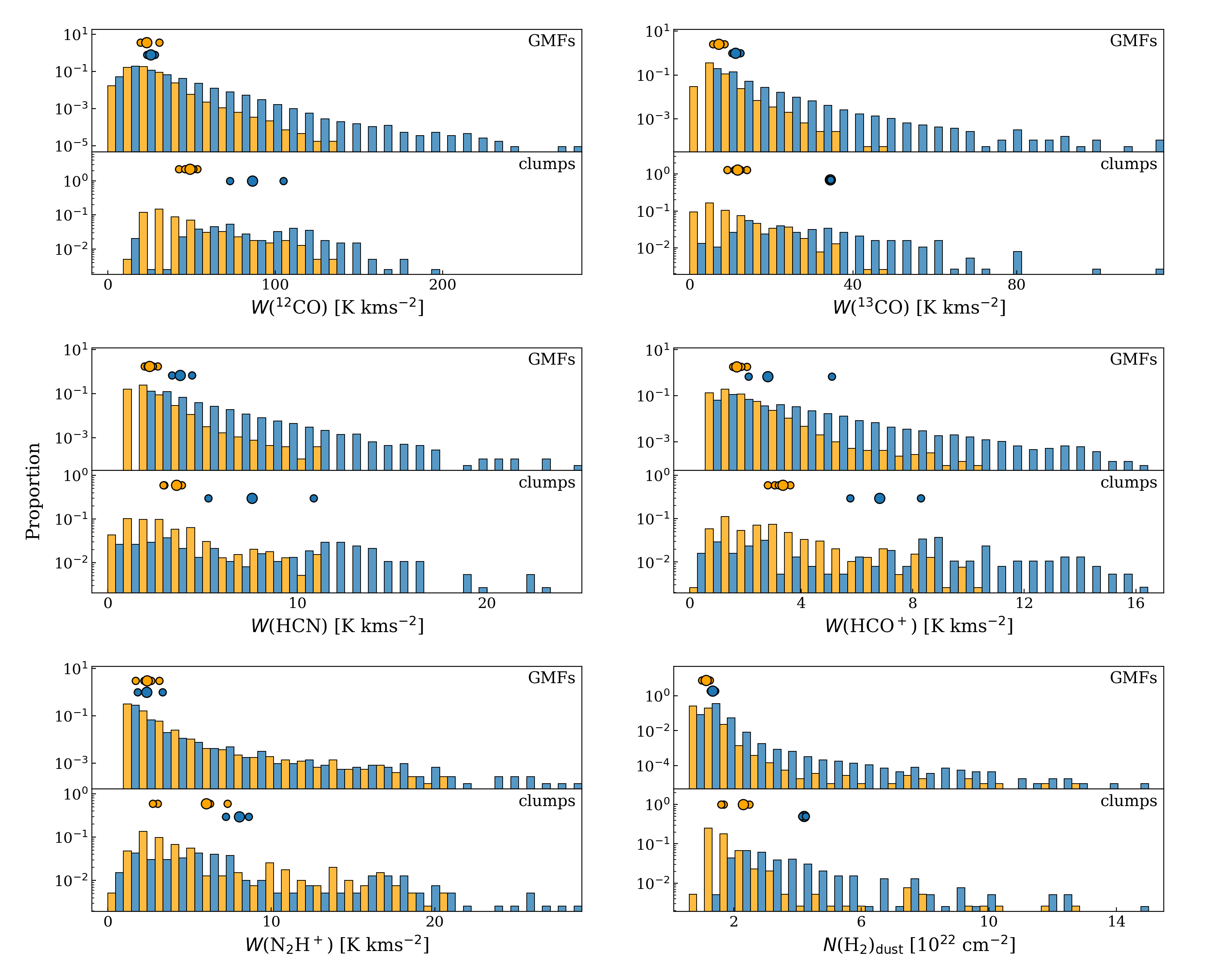}
    \caption{Comparison of the detected integrated intensity ranges of the different studied molecular species in the two GMFs. In each panel, the top sub-panel shows the distributions drawn from all pixels and the bottom sub-panel the distributions drawn from pixels of the N$_2$H$^+$-clumps (both only counting pixels with emission levels above 2\,$\times$\,$\sigma_{\mathrm{int}}$). The interarm distributions are coloured orange, the arm distributions blue. Above the histograms, large circles show the averages computed for the arm and interarm, respectively, while smaller circles indicate the averages for the individual arm and interarm observing regions. The vertical axis for each histogram is set so that the bar heights sum up to 1, and the scale is logarithmic due to the wide range shown by the values.}
    \label{fig:clumps_emissions_histo}
\end{figure*}

\begin{figure*}
    \centering
    \includegraphics[width=\linewidth]{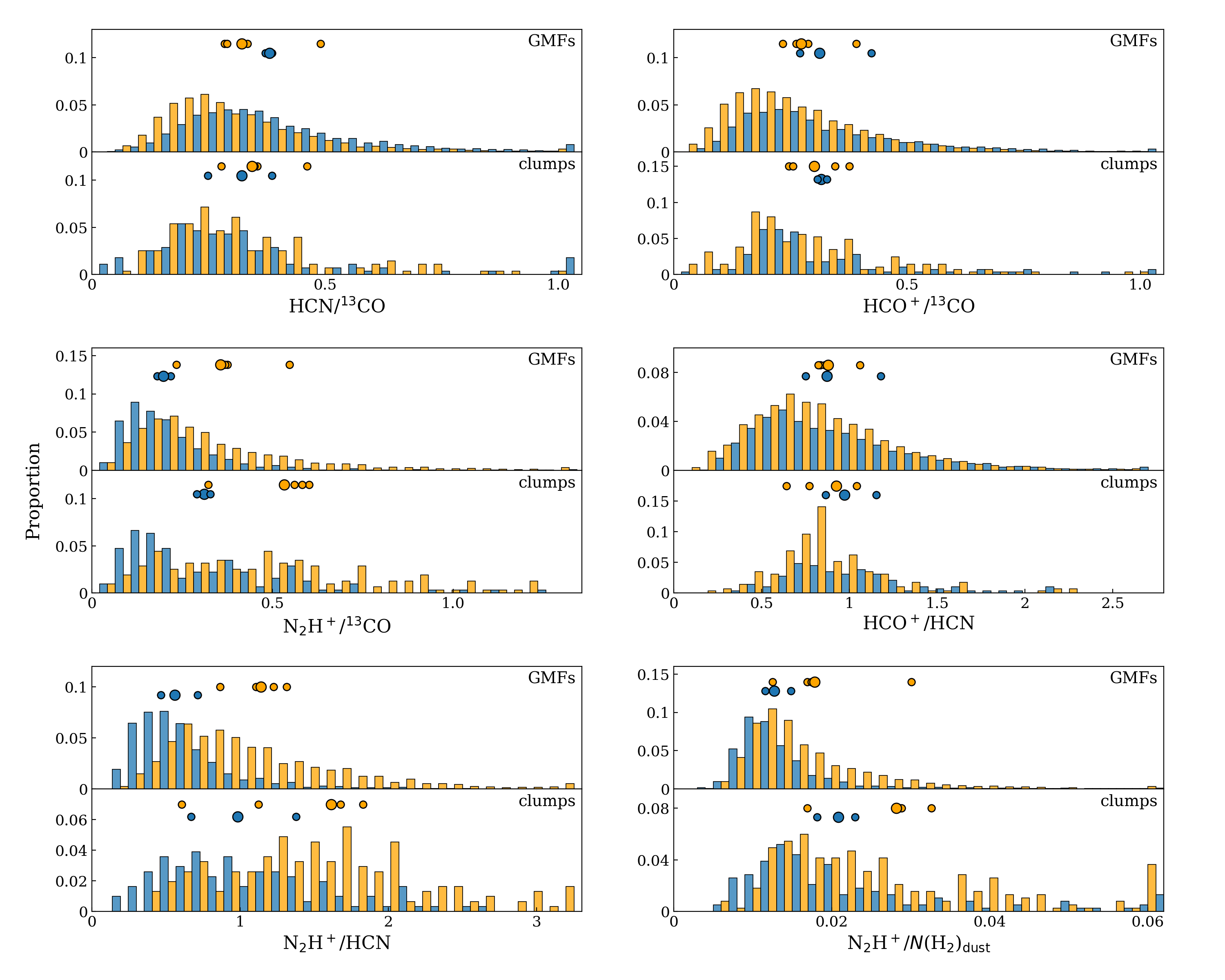}
    \caption{Comparison of the detected line ratios of the different studied molecular species in the two GMFs. The figure uses the same colors and markers as Fig.~\ref{fig:clumps_emissions_histo}, except here the y axis is linear.}
    \label{fig:clumps_ratios_histo}
\end{figure*}

\begin{table*}
    \centering
    \begin{tabular}{l | llll}
        \hline
        \multicolumn{1}{c}{Parameter} &  \multicolumn{1}{c}{KS2 GMF} & \multicolumn{1}{c}{KS2 clump}  & \multicolumn{1}{c}{KS2 arm} & \multicolumn{1}{c}{KS2 interarm} \\
        \hline 
        \hline
        $^{12}$CO          &   0.43 (6.5\,$\times$\,10$^{-3}$) & 0.17 (0.81) &  0.50 (9.0\,$\times$\,10$^{-4}$) & 0.30 (0.14) \\
        $^{13}$CO          &   0.60 (2.4\,$\times$\,10$^{-5}$) & 0.33 (0.07) &  0.33 (0.07)                     & 0.17 (0.81) \\
        HCN                &   0.40 (0.015) & 0.33 (0.07) &  0.47 (2.5\,$\times$\,10$^{-3}$) & 0.20 (0.59) \\
        HCO$^+$            &   0.47 (2.5\,$\times$\,10$^{-3}$) & 0.40 (0.02) &  0.66 (1.3\,$\times$\,10$^{-6}$ & 0.23 (0.39) \\ 
        N$_2$H$^+$         &   0.23 (0.39)                     & 0.23 (0.39) &  0.27 (0.24)                     & 0.23 (0.39) \\
        $N$(H$_2$)$_{\mathrm{dust}}$ &  0.43 (6.5\,$\times$\,10$^{-3}$)  &  0.23 (0.39) & 0.33 (0.07)           & 0.13 (0.96) \\
        \hline 
        HCN/$^{13}$CO      &     0.16 (0.84) & 0.09 (1.0) & 0.84 (8.3\,$\times$\,10$^{-12}$) & 0.69 (1.7\,$\times$\,10$^{-7}$) \\
        HCO$^+$/$^{13}$CO  &     0.19 (0.64) & 0.22 (0.43) & 0.75 (4.8\,$\times$\,10$^{-9}$)  & 0.59 (1.4\,$\times$\,10$^{-5}$) \\
        N$_2$H$^+$/$^{13}$CO   & 0.44 (0.004) & 0.25 (0.27) & 0.31 (0.09)       & 0.56 (5.2\,$\times$\,10$^{-5}$) \\
        HCO$^+$/HCN            & 0.16 (0.84) & 0.16 (0.84) & 0.72 (3.0\,$\times$\,10$^{-8}$)  & 0.66 (8.1\,$\times$\,10$^{-7}$) \\
        N$_2$H$^+$/HCN         & 0.41 (0.01) & 0.22 (0.43) & 0.38 (0.02)       & 0.63 (3.6\,$\times$\,10$^{-6}$) \\
        N$_2$H$^+$/$N$(H$_2$)$_{\mathrm{dust}}$  &  0.28 (0.16) & 0.22 (0.43) & 0.41 (0.01) & 0.59 (1.4\,$\times$\,10$^{-5}$)  \\
        \hline
    \end{tabular}
    \caption{Statistics (the maximum absolute difference between the empirical distribution functions of the samples) and p-values from the two-sample Kolmogorov-Smirnov tests. (1) The tested parameter; (2) Interarm data tested against arm data on the filament-scale; (3) Interarm data tested against arm data on the clump-scale; (3) Arm data on the filament- versus on the clump-scale; (4) Interarm data on the filament- versus on the clump-scale.}
    \label{tab:ks_test}
\end{table*}

We perform two-sample Kolmogorov-Smirnov (KS2) tests using the \textit{ks\_2samp} function of \textit{scipy} in \textit{python}. First we gather all the detected molecular emission and molecular ratio values on all pixels above 2\,$\times$\,$\sigma_{\mathrm{int}}$ in the two GMFs and call these the filament-scale distributions of line emissions and ratios. We also gather the molecular emission and ratio values detected \textit{only} inside the defined N$_2$H$^+$-clumps, again only on pixels above the relevant 2\,$\times$\,$\sigma_{\mathrm{int}}$ levels, and call these the clump-scale line emissions and ratios. We bin these sets of data, sorting the values into 30 bins between their minima and maxima, as seen on the histograms of Fig.~\ref{fig:clumps_emissions_histo} and \ref{fig:clumps_ratios_histo}. We note that the histograms are plotted in the same intervals as in Fig.~\ref{fig:allregion_emissions_box1} and \ref{fig:allregion_emissions_box2}, with outliers above the respective limits shown in the last histogram bins. The exception is $N$(H$_2$)$_{\mathrm{dust}}$ where we clipped the data at 1.5\,$\times$\,10$^{23}$\,cm$^{-2}$ in order to get a better resolution, since above this level only an insignificant amount of pixels show values up to 2.1\,$\times$\,10$^{23}$\,cm$^{-2}$. The results of the performed KS2 tests are summarized in Table~\ref{tab:ks_test}. 

For this type of statistical test, the underlying continuous distributions of two independent samples are compared. The null hypothesis is that the two distributions are identical. As a result of the test, we get a statistic and its p-value: the statistic is the maximum absolute difference between the empirical distribution functions of the two samples. It follows then that confirming that values in the two datasets tested against each other do not arise from the same underlying distribution (e.g. rejecting the null hypothesis) requires a p\,<\,0.05 if using a confidence level of 95\%.

The second column of Table~\ref{tab:ks_test} shows the statistics measured on the whole filament scale. Except for the N$_2$H$^+$ emission, all studied molecular integrated intensities and the $N$(H$_2$)$_{\mathrm{dust}}$ appear to arise from different underlying distributions, suggesting that the emission is environment-dependent. The N$_2$H$^+$ emission, on the other hand, appears the same in the arm and the interarm on filament scales (p\,=\,0.39).

The lower half of Table~\ref{tab:ks_test} shows how the line ratios compare between the arm and interarm. Our statistic shows that all ratios are drawn from the same underlying distribution, except for N$_2$H$^+$/$^{13}$CO (p\,=\,0.004) and N$_2$H$^+$/HCN (p\,=\,0.01). That these two ratios with N$_2$H$^+$ appear to be drawn from different distributions between the arm and interarm is what we would expect, given the behaviour of the $^{13}$CO and HCN emission discussed above. By that logic, we would also expect that the N$_2$H$^+$/$N$(H$_2$)$_{\mathrm{dust}}$ ratio would follow suit, but despite the different distributions shown by $N$(H$_2$)$_{\mathrm{dust}}$ between the arm and interarm, the ratio is consistent between the two.
 
The remaining molecular ratios (HCN/$^{13}$CO, HCO$^+$/$^{13}$CO, and HCO$^+$/HCN) showing the same distributions between arm and interarm environments is interesting, as even though the emission of these molecules individually appear environmentally dependent, the molecular ratios behave the same way in both environments. We conclude that on these whole filament scales (tens of parsecs), while the $^{13}$CO, HCN and HCO$^+$ molecular emission levels depend on which environment one probes, the ratios between them appear to be in lockstep. N$_2$H$^+$, however, does not track reliably with any of these molecules.

The third column of Table~\ref{tab:ks_test} shows the comparison of the clump-scale distributions of emissions in the arm and the interarm GMFs. Only the HCO$^+$ emission is proven to arise from different underlying distributions inside the clumps of the two environments, but the case of $^{13}$CO and HCN are also very close with p\,=\,0.07. The $^{12}$CO and N$_2$H$^+$ emissions seem to arise from the same distribution. The uniformity of the $^{12}$CO emission is not hugely surprising if the transition does not originate from the densest regions of the defined N$_2$H$^+$-clumps. For the molecular ratios (lower half of the table), with all p-values well above 0.05, the detected values reflect the same underlying distribution in the arm and the interarm. From the comparison between the arm and interarm at clump scales, we conclude that the arm and the interarm show different emission characteristics regarding HCO$^+$, HCN, and $^{13}$CO, however, they seem quite similar regarding the species tracing the lowest and the highest column densities ($^{12}$CO and N$_2$H$^+$). Clumps in the arm and in the interarm are also similar to each other regarding molecular ratios.

To check whether the behaviours are consistent between scales within the same environment, we perform a separate test. The fourth column of Table~\ref{tab:ks_test} compares the distributions of the detected emissions and line ratios measured on filament-scale against the distributions of the emissions and line ratios measured on clump-scale in the arm filament. Here, all molecular emissions except for N$_2$H$^+$ (and marginally, $^{13}$CO) seem to arise from different underlying distributions on the two scales. Since N$_2$H$^+$ mostly emits in the defined N$_2$H$^+$-clumps, the similarity of the two scales regarding this tracer is evident. Arm clumps seem to be distinct from their host filaments regarding most other tracers, however. We perform the same test for the interarm (fifth column). Here, the underlying distributions of the detected molecular emissions are similar on the two scales for all tracers, meaning that interarm clumps seem more similar to their host filaments than arm clumps. Regarding the molecular ratios, the resulting p-values range between 10$^{-12}-$10$^{-2}$, with the exception of N$_2$H$^+$/$^{13}$CO in the arm with a p\,=\,0.09, meaning that the underlying distributions for most studied molecular ratios are different for the interarm clumps from those of the entire filament.

\subsubsection{Star forming gas in the arm and interarm: an evolving view}

The use of multiple molecular tracers is a powerful tool to infer the conditions of star forming gas. We have explored possible environmental and scale dependencies in the above sections.
 
The most abundant tracer in our study, $^{12}$CO, shows the least variation in emission between the arm and the interarm, but we have shown that the $^{13}$CO(3$-$2) emission differs between these environments. The variation in the ratio of these two CO isotopologues computed for our two environments is in agreement with the level of variation claimed in \citet{rigby2024}. At the same time, HCN and HCO$^+$ also track changes in the $^{13}$CO emission and exhibit similar trends to each other between environments and spatial scales.
 
We find that N$_2$H$^+$ does not behave like other molecules. Its emission level is unchanged between environments, and as there is minimal extended emission, so it does not show scale-dependence either.

These results support the view that, while the moderate-density bulk of the gas does not show differences across kiloparsec-scale environments (no systematic variation in how $^{12}$CO traces H$_2$) a higher-density tracer like $^{13}$CO does show variations on large scales. Specifically, a variation of the CO emission between the arm and interarm, alongside similar changes seen in the HCN and HCO$^+$ emission, might be induced by the changing radiation field strength connected to local star formation \citep[e.g.][]{clark2015}. If CO survives more effectively inside molecular clouds and clumps in the arm, since it is better shielded (higher column densities due to the concentration of material), the lower-density tracing $^{12}$CO might get excited and destroyed more uniformly, while $^{13}$CO, emitting from higher-density material that is more sensitive to shielding can increase its filling factor inside clumps and appear as brighter depending on the spatial resolution of our observations. Additionally, its emission can be boosted by the heating of the gas by ongoing star formation. If the observed interarm clouds are less well-shielded in more diffuse gas against UV photons arriving from internal star formation (or even high-mass star forming regions at the edge of a neighbouring spiral arm), the CO brightness will decrease, even if heating by star formation and/or cosmic rays would increase its emission. In this picture, an even higher-density tracer, N$_2$H$^+$, only starts emitting in the centre of the densest, well-shielded clumps, and its emission characteristics will reflect the immediate, local clump environment. This explains its uniformity between different large-scale environments while N$_2$H$^+$ can still be potentially enhanced or destroyed by clump interior processes locally. In other words, once dense clumps form and are detectable in N$_2$H$^+$ emission, their properties are independent of the environment in which they formed. Large-scale environment might influence the time scales of clump formation or the crowding of sources, but N$_2$H$^+$-traced clumps (star-forming or on the verge of it) are more detached from their environments \citep{kauffmann2017, pety2017, priestley2023b}. 

\subsection{Star formation in the dense, N$_2$H$^+$-traced clumps in the two GMFs}

We now consider the concentration of truly dense gas inside the identified dense clumps in our two GMFs and the star formation connected to them. 

The global filling factors we calculated in Section~\ref{sec:covandsum} are dependent on the mapped area. In an attempt to give a more reliable measure of dense gas coverage, we use our N$_2$H$^+$-clumps to derive a parameter we call clump fraction. We define the clump fraction as the ratio of all pixels located in N$_2$H$^+$-clumps versus all pixels inside the 2\,$\times$\,$\sigma_{\mathrm{int}}$ contour of $W$($^{13}$CO) which are largely well-defined, largely closed-contour areas on our $^{13}$CO integrated intensity maps. The parameter $\sigma_{\mathrm{int}}$ here is the average of the individual $\sigma_{\mathrm{int}}$ values measured on the individual observing areas, and its value is 2.4\,K\,km s$^{-1}$. There are 22 N$_2$H$^+$-clumps in the interarm and 16 in the arm, all with similar angular sizes, possible to approximate with ellipses with average major axes of 33\,arcsec and average minor axes of 20.2\,arcsec \citep{feher2024}. The phyiscal size of the area enclosed by the 2\,$\times$\,$\sigma_{\mathrm{int}}$ contour on the $W$($^{13}$CO) map is 316.6\,pc$^2$ in the interarm and 132.2\,pc$^2$ in the arm. At the same time, the total area covered by clumps in the interarm is around 4.4 times than what is covered in the arm (8.3\,pc$^2$ and 1.9\,pc$^2$). This then results in clump fraction values of 2.6\% for the interarm and 1.5\% for the arm. The difference is even larger if we use the 3\,$\sigma_{\mathrm{int}}$ contour, which is seen as a more strictly closed contour on all observing regions, resulting in 7.0\% for the interarm and 2.8\% for the arm. We conclude that dense, N$_2$H$^+$-traced clumps encompass very small areas in both GMFs, and this area is significantly smaller for the arm than the interarm, even compensating for the different physical scales measured. This result is interesting, but to speculate on the underlying effects requires more observational examples. We note however that beam dilution towards the more distant cloud, the interarm GMF, might make the detected clumps appear larger there.

\begin{figure}
    \centering
    \includegraphics[width=0.85\linewidth]{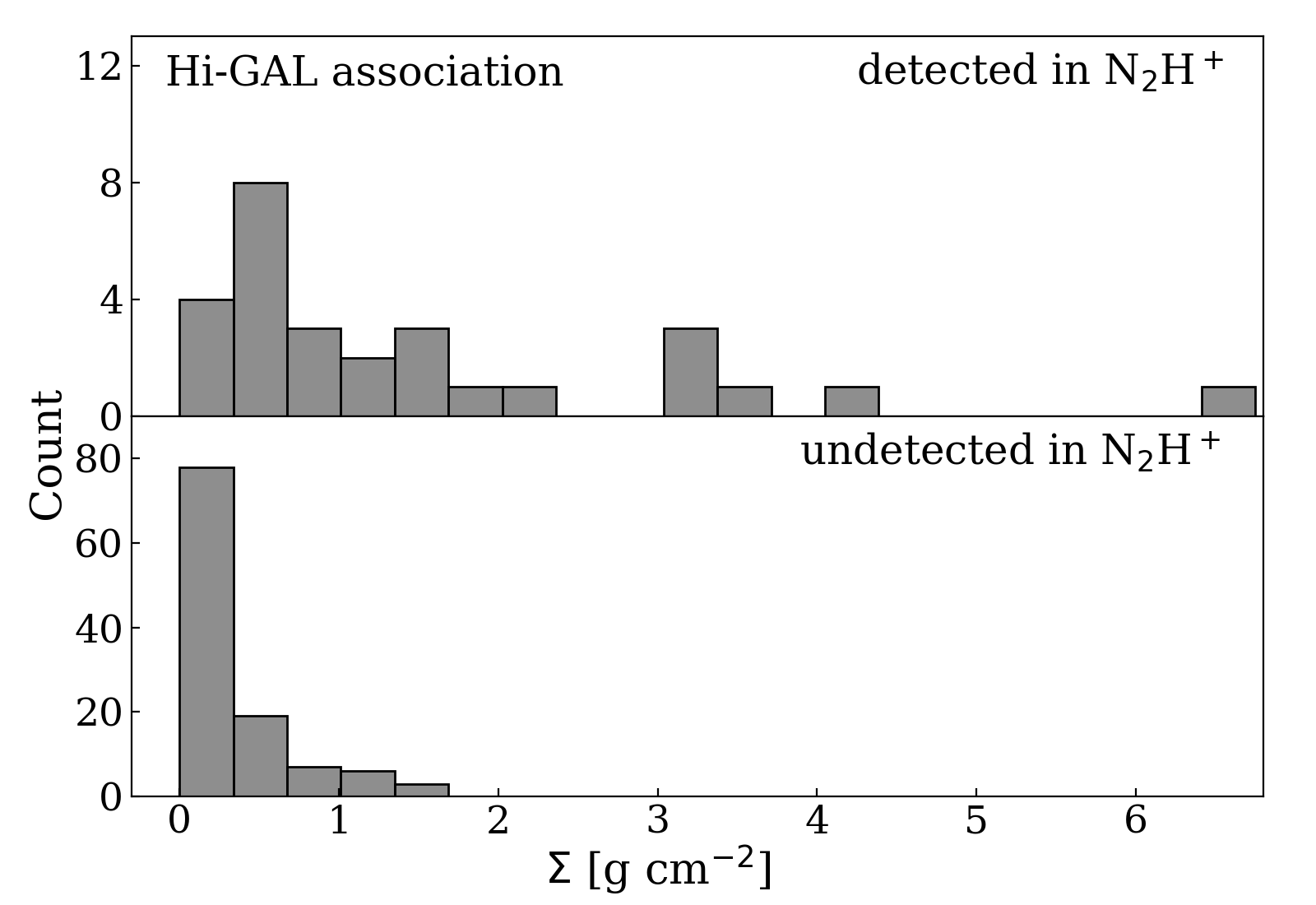}
    \includegraphics[width=0.85\linewidth]{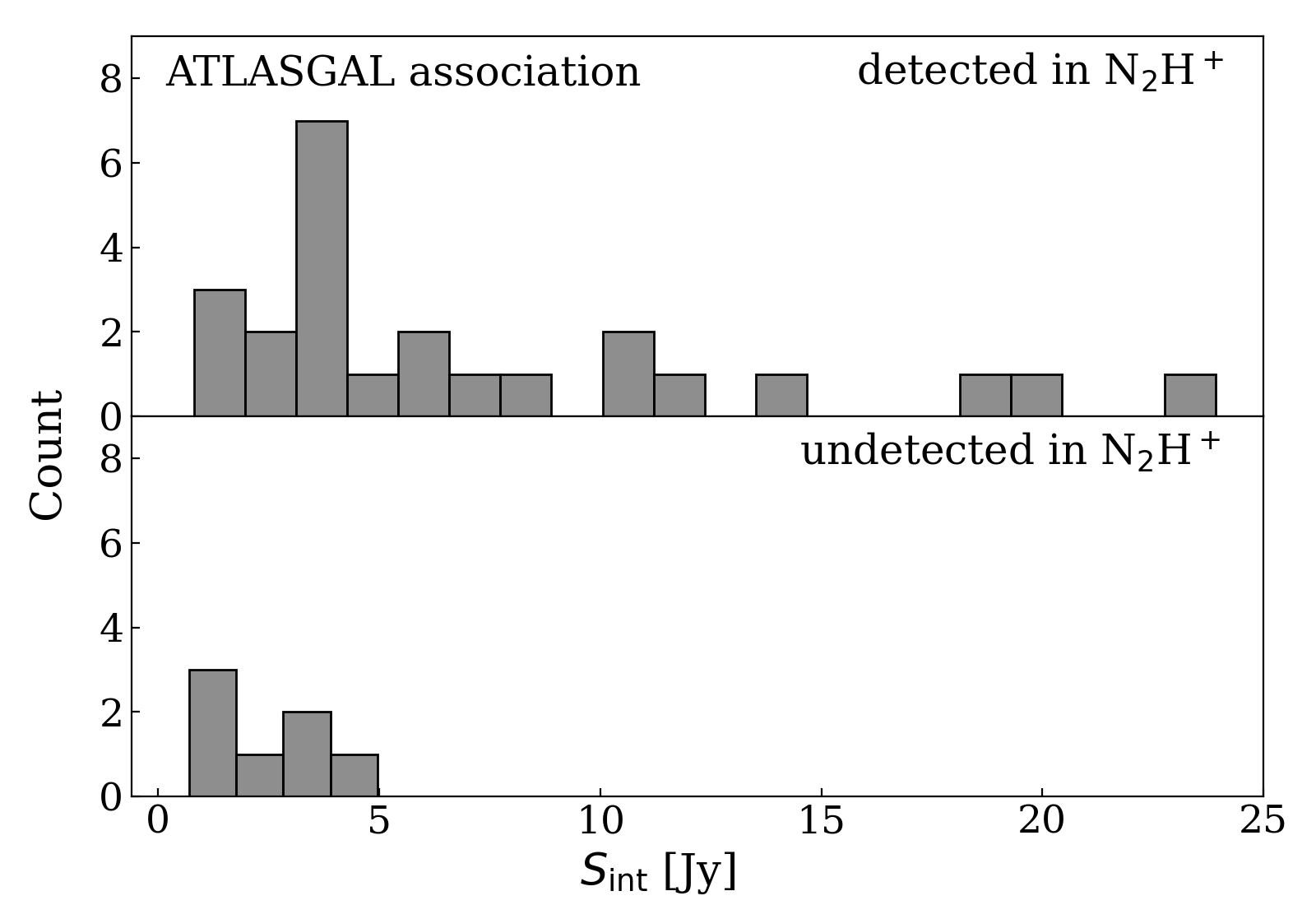}
    \caption{Comparison of the distributions of clumps detected and undetected by our N$_2$H$^+$ observations in the Hi-GAL and the ATLASGAL compact source catalogues.}
    \label{fig:atlashigal}
\end{figure}

In \citet{feher2024} we compared our list of N$_2$H$^+$-clumps to the objects listed in the Hi-GAL compact source catalogue. We perform a similar association here to evaluate how we detect dense gas with different methods and what consequences it brings to our estimation of global star formation rates (SFR). We use the same Hi-GAL 360$^{\circ}$ catalogue by \citet{elia2021}, and additionally, the ATLASGAL compact source catalogue by \citet{contreras2013}, searching in a circle of 20$\arcsec$ radius around the N$_2$H$^+$-clump central coordinates to find associated structures. Of the 38 N$_2$H$^+$-clumps 28 are detected in the Hi-GAL catalogue (13 in the interarm, 15 in the arm) and 24 in the ATLASGAL catalogue (11 in the interarm, 13 in the arm). 21 N$_2$H$^+$-clumps are detected by both (9 in the interarm, 12 in the arm). The distribution of N$_2$H$^+$-clumps that are not detected by these continuum-based methods generally peak towards lower $^{12}$CO and $^{13}$CO integrated intensities and exhibit lower $N$(H$_2$)$_{\mathrm{dust}}$ than those detected. The difference regarding HCN, HCO$^+$ and N$_2$H$^+$ is not that significant. From the total number of 141 Hi-GAL clumps found on the mapped area of our IRAM observations only 28 are also N$_2$H$^+$-clumps (19.8\%); the total number of ATLASGAL objects in our observing area is 31 from which 24 (77.4\%) are detected as N$_2$H$^+$-clumps. The clumps in the Hi-GAL and the ATLASGAL catalogues that were not detected by our N$_2$H$^+$ observations and subsequent dendrogram source extraction generally show lower surface densities and integrated fluxes at 870\,$\mu$m as these are given in the Hi-GAL and ATLASGAL catalogues, respectively (see Fig.~\ref{fig:atlashigal}). The Hi-GAL survey was more sensitive than ATLASGAL, and able to detect more fainter sources, accounting for the higher source number on our regions, while the ATLASGAL list of dense clumps on the whole correspond better with our N$_2$H$^+$-detected clumps. 

The discrepancy between what is detected as dense gas by the Hi-GAL catalogue and our N$_2$H$^+$-survey towards the GMFs may bring consequences to our assumptions about the estimated SFR. To highlight this, we determine the global SFR for the two GMFs applying the method of \citet{elia2022} (Equation 1), based on the work by \citet{veneziani2017}, first using the object masses from the full list of Hi-GAL clumps located on our observing area, then only the masses of those clumps that were also detected in N$_2$H$^+$. The SFR derived with only clumps detected by N$_2$H$^+$ is 22\% of the SFR derived using all Hi-GAL clumps in the two GMFs, thus the Hi-GAL-derived SFR is more than four times as high as what we would get only considering the N$_2$H$^+$-traced clump structures. In both cases, the SFR of the arm is around 65\% of the total SFR. This means that even though the N$_2$H$^+$ emission is concentrated into a smaller fraction of the moderate-density gas (i.e. smaller clump fraction, see above) in the arm, the estimated star forming activity of the arm is higher.

The result above is a stark demonstration that the choice of dense gas tracer, especially in unresolved (extragalactic) investigations of star formation, can dominate systematics and scatter in inferring the star formation rate. If N$_2$H$^+$(1$-$0) is indeed the most reliable tracer of dense star forming gas on clump scales \citep{priestley2023b}, then use of alternative tracers such as dust emission can significantly overestimate the amount of truly star-forming gas.

\subsection{Mid-scale spatial variation of the clump molecular ratios along the two GMFs}
\label{sed:sub:spatial}

We have seen that the filament-scale average N$_2$H$^+$/HCN ratio is 0.6\,$\pm$\,0.3 in the arm GMF and 1.1\,$\pm$\,0.6 in the interarm, a significant difference. There is a similar preference towards lower and higher values, respectively, when only considering values measured in the arm and interarm clumps. At the same time, the KS2 tests showed that dense clumps exhibit the same kind of underlying distributions in the arm and the interarm regarding the N$_2$H$^+$/HCN ratio and most of the studied line ratios: there does not seem to be a significant difference between clumps in different environments when considering line ratios. On the other hand, the HCO$^+$/HCN ratio is remarkably similar in the GMFs on both the filament-scale and the clump-scale, but we can still see variations in between observing regions. There are clearly many contributing factors to the spatial variations of these two ratios on different scales and in different environments. We further analyse this in the following. 

\begin{figure*}
    \centering
    \includegraphics[width=\linewidth]{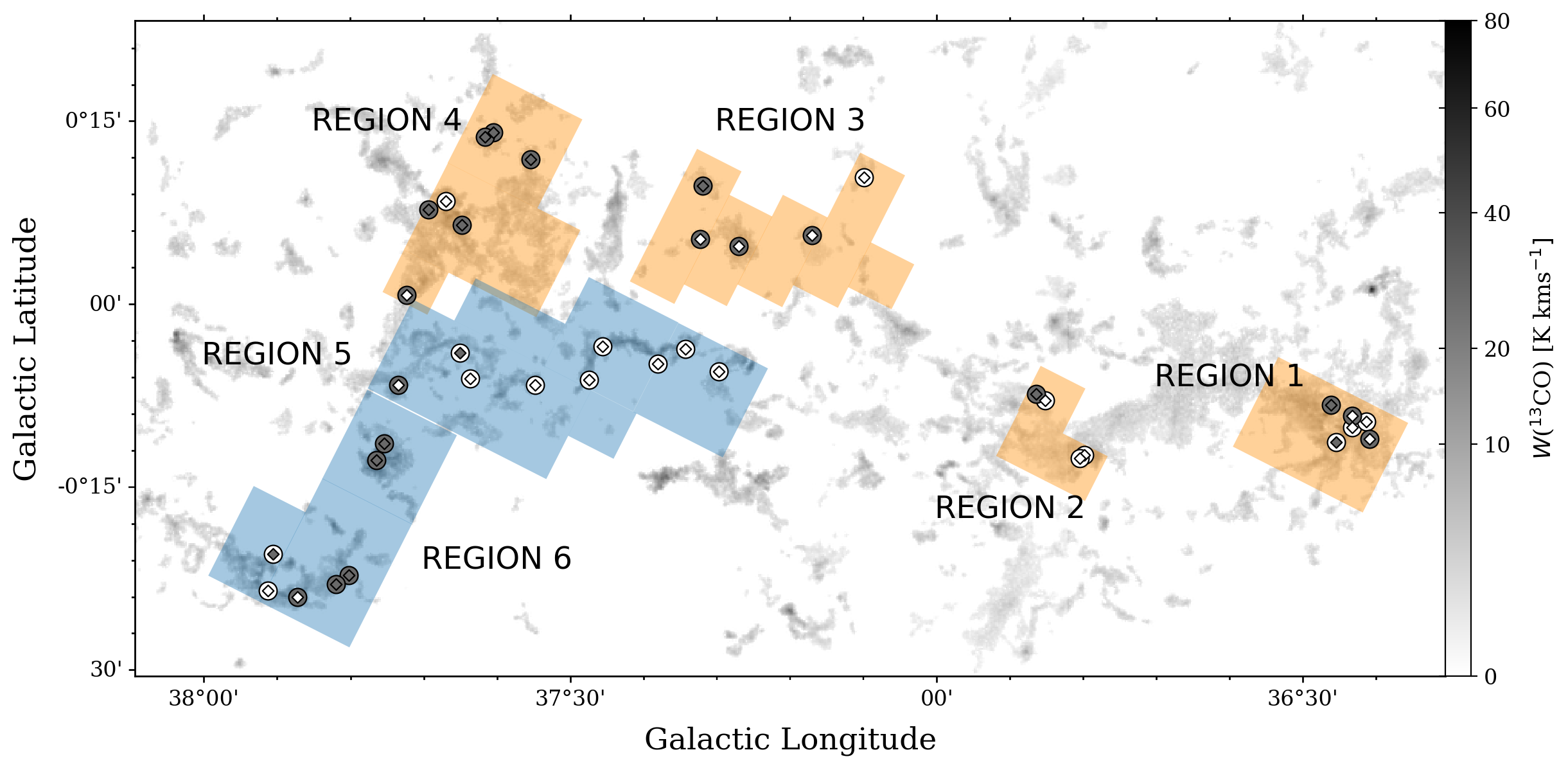}
    \caption{The variation of the N$_2$H$^+$/HCN and HCO$^+$/HCN ratios inside the N$_2$H$^+$-clumps along the two GMFs. The underlying color map is the same as in the bottom panel of Fig.~\ref{fig:overview}. Clumps with N$_2$H$^+$/HCN\,>\,1 are marked with dark grey circles, and clumps with N$_2$H$^+$/HCN\,<\,1 are marked with white circles. Clumps with HCO$^+$/HCN\,>\,1 are also marked with dark grey diamonds, and clumps with HCO$^+$/HCN\,<\,1 are marked with white diamonds.}
    \label{fig:cluster_ratio_map}
\end{figure*}

\begin{figure}
    \centering
    \includegraphics[width=\linewidth]{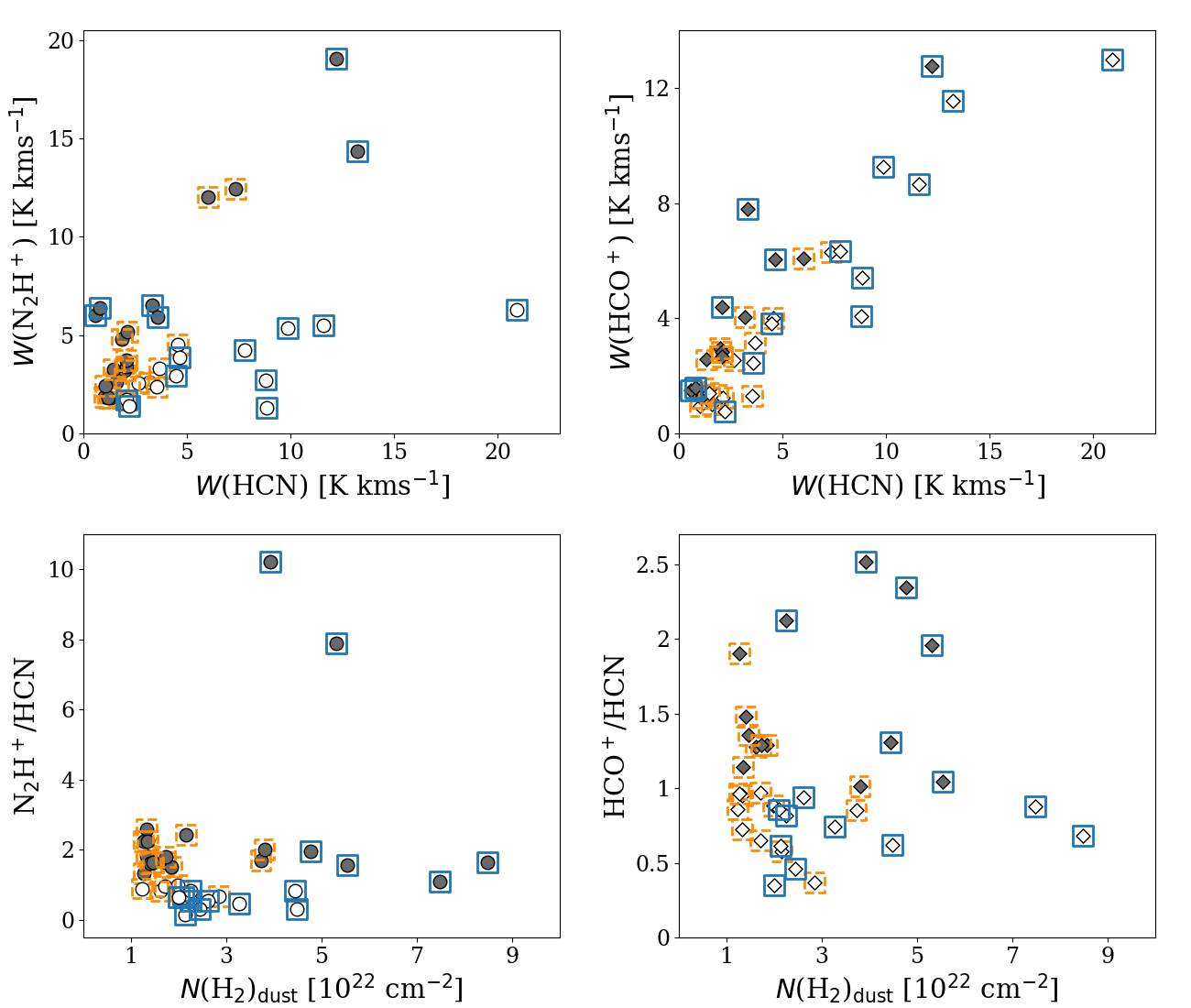}
    \caption{Correlations of the N$_2$H$^+$/HCN and HCO$^+$/HCN ratio with HCN, HCO$^+$, and N$_2$H$^+$ emission, and with $N$(H$_2$)$_{\mathrm{dust}}$ in the N$_2$H$^+$-clumps. The ratios are marked the same way and in Fig.~\ref{fig:cluster_ratio_map}. Orange dashed square outlines the markers that indicate interarm clumps, and blue solid square outlines the markers that indicate arm clumps.}
    \label{fig:cluster_ratio_corr}
\end{figure}

We defined the average N$_2$H$^+$/HCN or the HCO$^+$/HCN ratio of a N$_2$H$^+$-clump ``low'' if it is below and ``high'' if it is above unity. Fig.~\ref{fig:cluster_ratio_map} shows the location of the N$_2$H$^+$-clumps and their N$_2$H$^+$/HCN and HCO$^+$/HCN ratios, both colour-coded to show low and high values. Fig.~\ref{fig:cluster_ratio_corr} also shows the relations of the two ratios with HCN, HCO$^+$, and N$_2$H$^+$ emission and column density.

It is immediately apparent that inside Region\,5 (the W-E orientated part of the arm GMF) 7 out of 9 N$_2$H$^+$-clumps show both low N$_2$H$^+$/HCN and low HCO$^+$/HCN ratios, while, in contrast, inside Region\,6 (the N-S orientated part) both ratios are high in 4 out of 7 clumps. This increase of the N$_2$H$^+$/HCN ratio from one part of the filament to the other correlates with the increase of the N$_2$H$^+$ emission in the clumps. The arm clumps that show high N$_2$H$^+$/HCN ratios also tend to have higher column densities. Two clumps in Region\,6 show especially high N$_2$H$^+$/HCN ratios, around 7.9 and 10.2, these are R6CL3 and R6CL4 \citep[see][]{feher2024}: cold, with average N$_2$H$^+$ emission levels but exhibiting barely any HCN emission. Regarding the HCO$^+$/HCN ratio, most arm clumps with higher HCO$^+$/HCN values are so because of low HCN emission. In contrast with the N$_2$H$^+$/HCN ratio, the variation of HCO$^+$/HCN has no correlation with column density. In chemical models, HCO$^+$ is less affected by freeze-out than HCN, thus changes in the HCO$^+$/HCN ratio between two regions with the same column density might indicate different volume densities or different temperatures, i.e. in the region with higher volume densities, HCN depletes more, thus the ratio increases \citep{priestley2023a, panessa2023}.

In the interarm GMF, the picture is more confused, but regularities can still be found. Regions\,1 and 2 (the western end of the GMF) contain more clumps with low N$_2$H$^+$/HCN ratios than Region\,3 and 4 (the eastern end). Generally, Region\,3 has clumps with low HCO$^+$/HCN ratios, while the Region\,4 clumps show high ratios, but in contrast with the arm filament, neither ratio show a correlation with column density. The increase of the N$_2$H$^+$/HCN ratio tends to be governed by the lower HCN emission in most clumps here.

Both filaments thus show local regularities in the average molecular ratios of their dense clumps, possibly governed by larger than clump-scale but smaller than filament-scale factors. These kinds of variations will also have an impact on our conclusions when analysing ten-parsec to kiloparsec-scale observations of the spiral arms or interarm regions of external galaxies.

\section{Conclusions} 
\label{sec:concl}

We present molecular emission mapping of the J\,=\,1$-$0 transitions of HCN, HCO$^+$, and N$_2$H$^+$, alongside the J\,=\,3$-$2 transitions of $^{12}$CO and $^{13}$CO over 500 square arcminutes area in the Galactic Plane centred near $l$\,$\sim$\,37$^{\circ}$ towards two Giant Molecular Filaments, one associated with the Sagittarius spiral arm, the other with an interarm structure. We assess and compare the emission properties across the two different environments from parsec-scale dense clumps traced by N$_2$H$^+$ to giant filament scale, and relate our observations to similar Galactic results and for ten-parsec-scale clouds now becoming resolvable in nearby galaxies. Our key results are as follows:

\begin{enumerate}
    \item Among the studied species, N$_2$H$^+$(1$-$0) has the lowest global filling factor on the mapped area, and lower in the arm than in the interarm (1.5\% and 7.1\%, respectively, of roughly equal angular areas). The HCO$^+$(1$-$0) and the HCN(1$-$0) emission show similar filling factors with 12$-$18\%, and $^{12}$CO(3$-$2) has the highest, above 85\%. The cumulative filling factors as a function of H$_2$ column density show a lot of variation along the interarm GMF for most studied species, while this is not characteristic of the arm GMF.
   \item Based on cumulative fractions, N$_2$H$^+$ traces high column density structures best among the studied species, and better in the arm than in the interarm. The curves for both GMFs resemble the ones derived for Orion\,A and W49 (a low- and a high-mass galactic star forming region), however, the arm GMF is more similar to these regarding the more distinct behaviour of the species and the traced $N$(H$_2$)$_{\mathrm{dust}}$ range. The interarm GMF is more likely to have detectable N$_2$H$^+$(1$-$0) emission at lower $N$(H$_2$)$_{\mathrm{dust}}$ compared to the arm.
   \item The global properties of $^{12}$CO(3$-$2) and N$_2$H$^+$(1$-$0) emission do not vary significantly between the arm and interarm environments that we have defined, while the arm filament is brighter in $^{13}$CO(3$-$2), HCN(1$-$0), and HCO$^+$(1$-$0) emission. Variations in the emission of $^{12}$CO(3$-$2), HCO$^+$(1$-$0), and N$_2$H$^+$(1$-$0) are often greater between regions of the same environment than between different environments.
   \item Considering the densest, N$_2$H$^+$-traced clumps, we measure tight correlations between all their detected molecular emission, and between H$_2$ column densities and molecular emission. The correlations of line emission with column densities are stronger for clumps in the arm GMF than for interarm clumps. An especially strong correlation stands between the $N$(H$_2$)$_{\mathrm{dust}}$ and the N$_2$H$^+$ integrated intensities of arm clumps, while the same relation for the interarm clumps is weak.
\end{enumerate}
Given these results, we looked in more detail at the integrated intensity ratios between our studied species, i.e. the molecular ratios HCN/$^{13}$CO, HCO$^+$/$^{13}$CO, N$_2$H$^+$/$^{13}$CO, HCO$^+$/HCN, N$_2$H$^+$/HCN, and N$_2$H$^+$/$N$(H$_2$)$_{\mathrm{dust}}$.
\begin{enumerate}
    \item The N$_2$H$^+$/$^{13}$CO and N$_2$H$^+$/HCN ratios, alongside N$_2$H$^+$/$N$(H$_2$)$_{\mathrm{dust}}$, show large variations between the two GMFs and inside them. These changes are governed by the different $^{13}$CO, HCN, and $N$(H$_2$)$_{\mathrm{dust}}$ levels, respectively, not N$_2$H$^+$. The variation of N$_2$H$^+$/$N$(H$_2$)$_{\mathrm{dust}}$ as a function of $N$(H$_2$)$_{\mathrm{dust}}$, especially that of the arm filament, is similar to the trends of Orion\,A and W49 in the relevant $N$(H$_2$)$_{\mathrm{dust}}$ range. 
   \item In contrast, the HCO$^+$/HCN ratio is remarkably similar between the arm and the interarm GMF, even though there are variations of it within the two structures.
   \item Two-population statistical tests between the arm and the interarm, then between the global filaments and the emission of dense clumps imply that the two GMFs are significantly different globally in most studied tracers, but their N$_2$H$^+$ emission characteristics are similar. Dense clumps of the arm appear different from their host filament in most studied line emission, while interarm clumps less so. The detected N$_2$H$^+$ emission is the most uniform between the two GMFs and filament- and clump-scales among the studied species.
   \item Even though dense clumps encompass a smaller fraction of the moderate-density gas (i.e. traced by $^{13}$CO) in the arm than in the interarm, we estimate a higher star-formation rate in the arm. At the same time, these estimated global star-formation rates are much lower when relying on N$_2$H$^+$ to mark the truly dense, star-forming gas, rather than on the far-infrared/sub-millimeter dust emission.
\end{enumerate}

We see that in the targeted GMFs, variations in the DGT emission are often more significant within the same environment than between the arm and the interarm. At the same time, the filling factors of the N$_2$H$^+$-emitting regions are so small that current and near-future extragalactic observations are not able to resolve them well. The N$_2$H$^+$/HCN ratio is clearly not constant on clump-scales, but spatial variations may average out when hundreds of clumps are integrated into one beam, leaving only changes that reflect large-scale differences in the environment or imply a constant value with a large scatter. Both large- and mid-scale effects will influence the line emission properties we detect. All these factors also bring consequences on estimating the global SFRs for large Galactic regions and extragalactic star forming clouds.

\section*{Acknowledgements}

The authors (OF, SER, FDP, and PCC) acknowledge the support of a consolidated grant (ST/W000830/1) from the UK Science and Technology Facilities Council (STFC). This work is based on observations carried out under project number 033-17 and E02-22 with the IRAM 30m telescope. IRAM is supported by INSU/CNRS (France), MPG (Germany) and IGN (Spain). The James Clerk Maxwell Telescope has historically been operated by the Joint Astronomy Centre on behalf of the Science and Technology Facilities Council of the United Kingdom, the National Research Council of Canada and the Netherlands Organisation for Scientific Research. This research made use of Astropy \citep[][https:// astropy.org]{astropy2022}, a community-developed core Python package for Astronomy as well as the Python packages NumPy \citep[][https://numpy.org]{harris2020array}, SciPy \citep[][https://scipy.org]{scipy2020}, and Matplotlib \citep[][https://matplotlib.org]{hunter2007}. This publication has also made use SAOImageDS9 \citep[][http://ds9.si.edu]{joye2003}, an astronomical imaging and data-visualization application and the CLASS and GREG software packages of GILDAS (https://www.iram.fr/IRAMFR/GILDAS).

\section*{Data Availability}

The data underlying this article are available in the Research Data Repository of Cardiff University, at https://dx.doi.org/10.17035/cardiff.29268926.



\bibliographystyle{mnras}
\bibliography{feher_iram_paper2_bib} 




\appendix

\section{Line ratio maps}
\label{app:intint}

\begin{figure*}
    \centering
    \includegraphics[width=\linewidth]{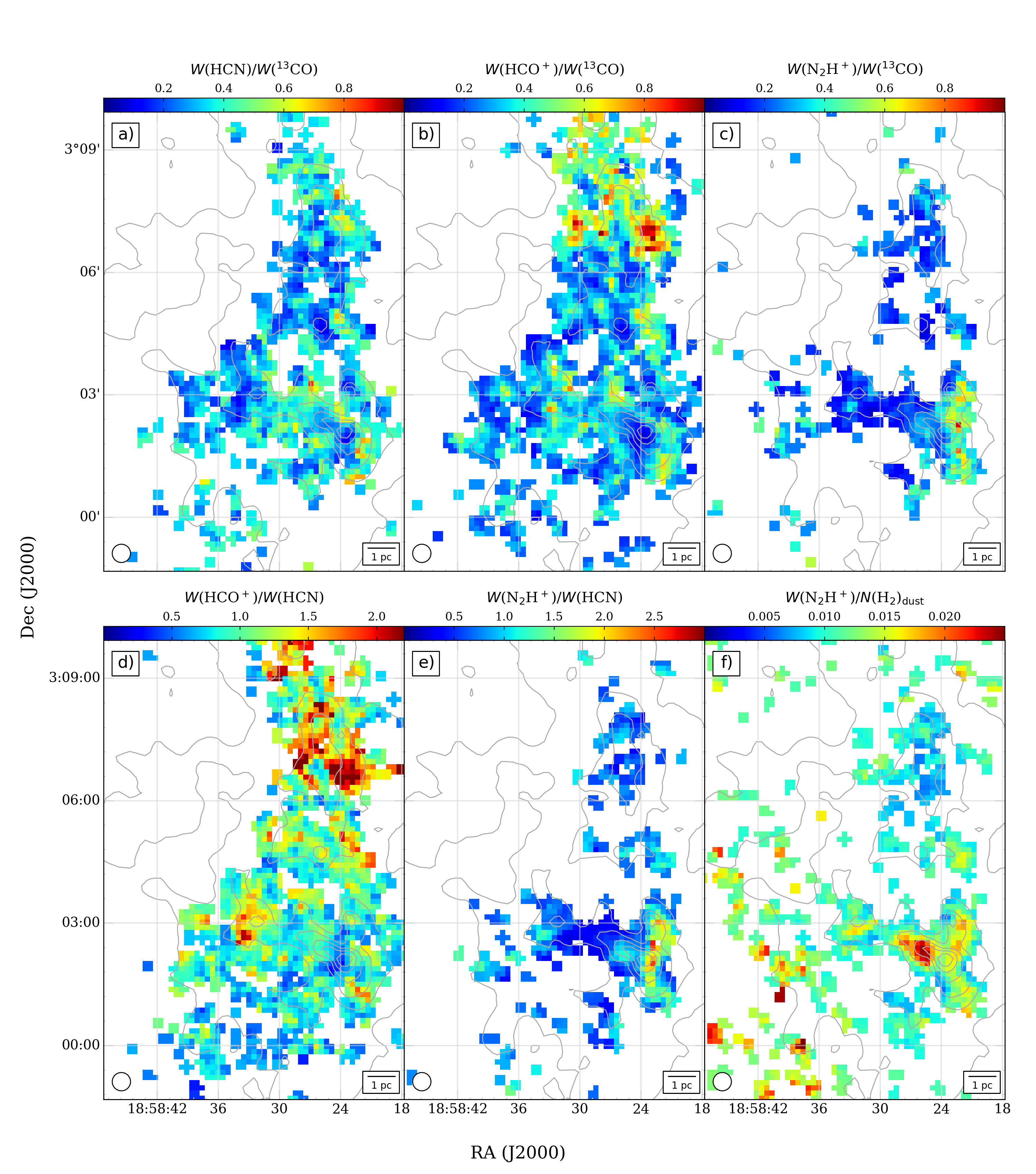}
    \caption{Molecular ratios in Region\,1: a) $W$(HCN)/$W$($^{13}$CO), b) $W$(HCO$^+$)/$W$($^{13}$CO), c) $W$(N$_2$H$^+$)/$W$($^{13}$CO), d) $W$(HCO$^+$)/$W$(HCN), e) $W$(N$_2$H$^+$)/$W$(HCN), f) $W$(N$_2$H$^+$)/$N$(H$_2$)$_{\mathrm{dust}}$. The gray contours mark the $N$(H$_2$)$_{\mathrm{dust}}$ H$_2$ column densities at the same levels as in Fig.~\ref{fig:reg1_morph}.}
    \label{fig:reg1_ratio}
\end{figure*}

\begin{figure*}
    \centering
    \includegraphics[width=\linewidth]{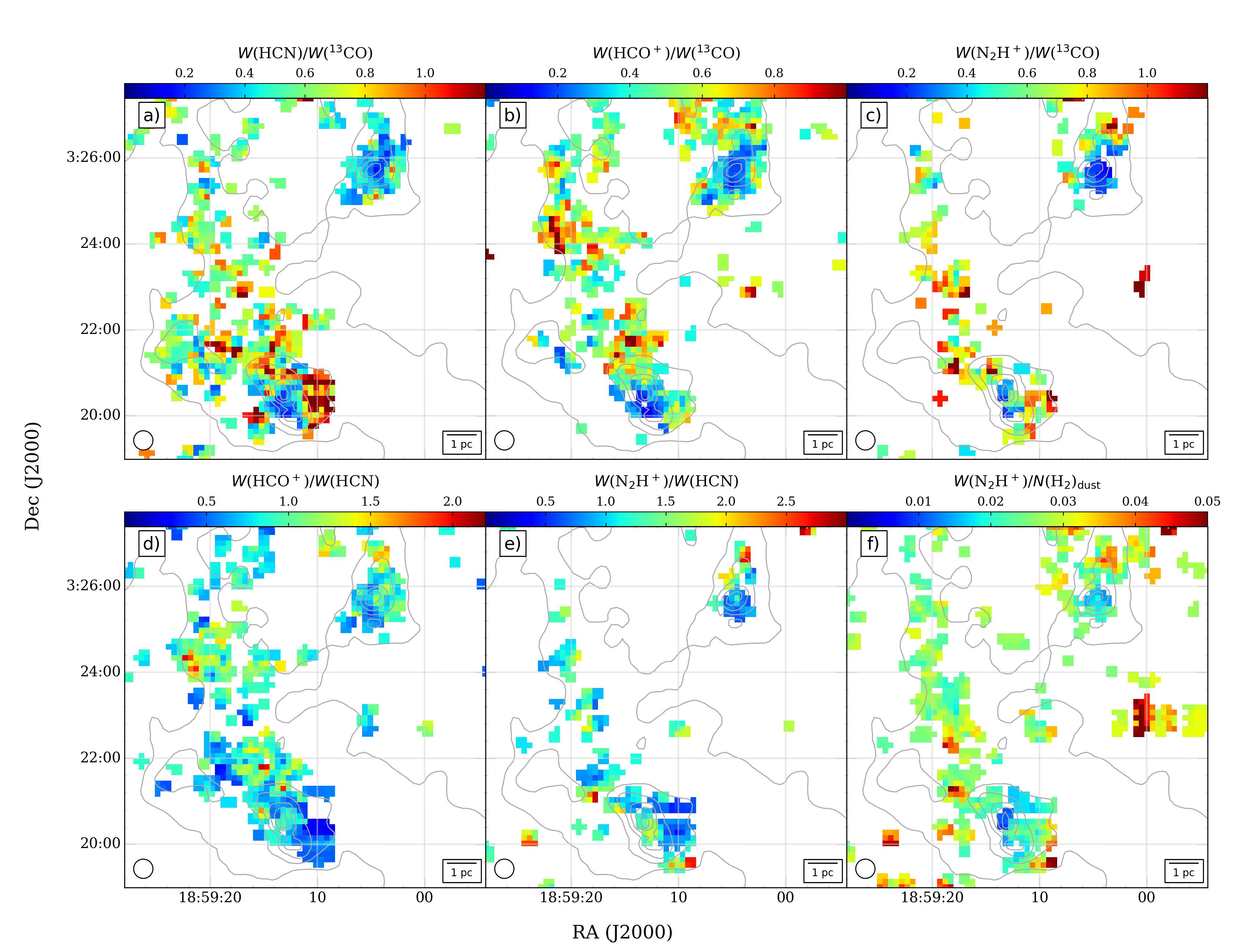}
    \caption{The same maps for Region\,2 as in Fig.~\ref{fig:reg1_ratio}. The gray contours mark the $N$(H$_2$)$_{\mathrm{dust}}$ H$_2$ column densities at the same levels as in Fig.~\ref{fig:reg2_morph}.}
    \label{fig:reg2_ratio}
\end{figure*}

\begin{figure*}
    \centering
    \includegraphics[width=\linewidth]{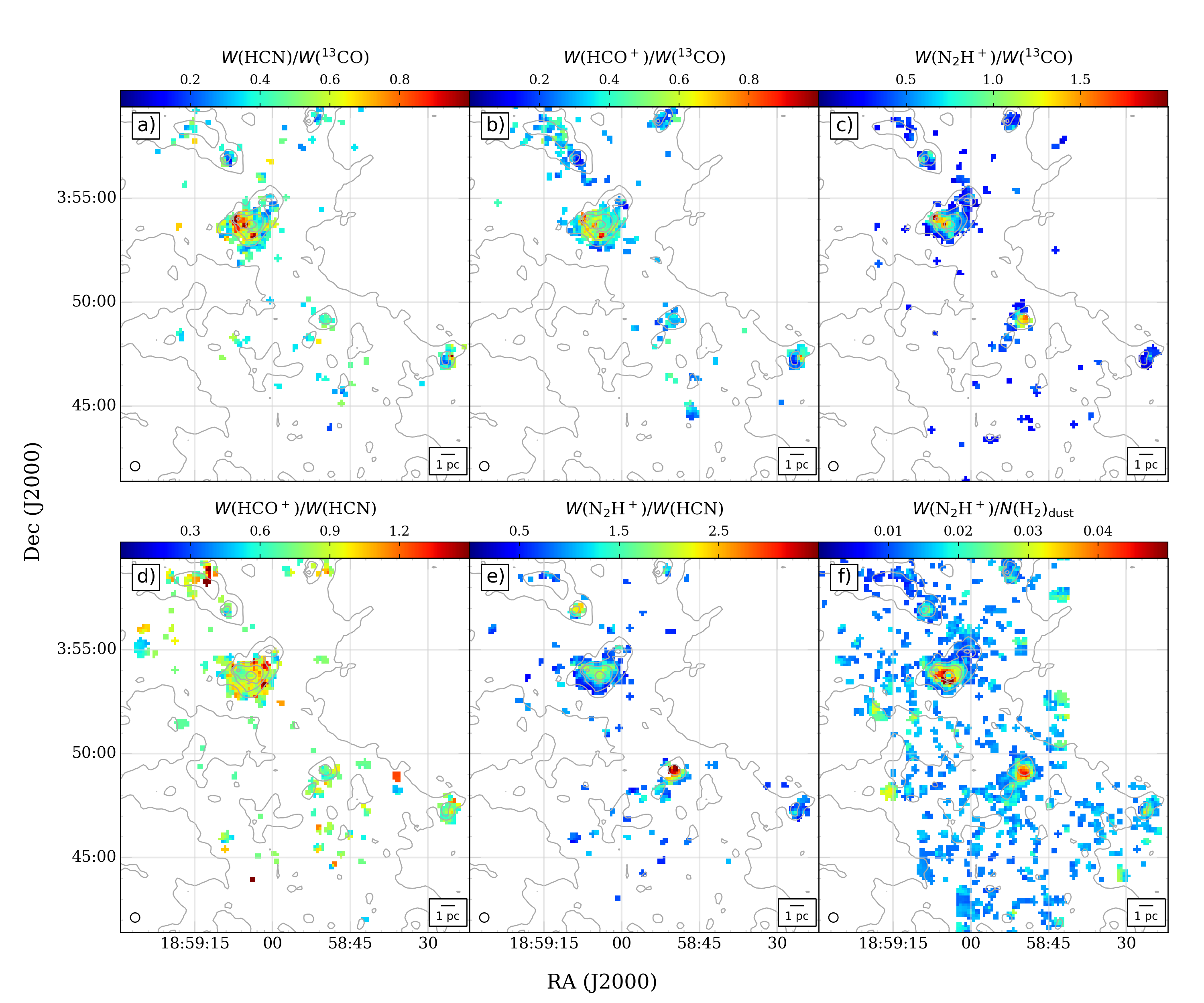}
    \caption{The same maps for Region\,3 as in Fig.~\ref{fig:reg1_ratio}. The gray contours mark the $N$(H$_2$)$_{\mathrm{dust}}$ H$_2$ column densities at the same levels as in Fig.~\ref{fig:reg3_morph}.}
    \label{fig:reg3_ratio}
\end{figure*}

\begin{figure*}
    \centering
    \includegraphics[width=\linewidth]{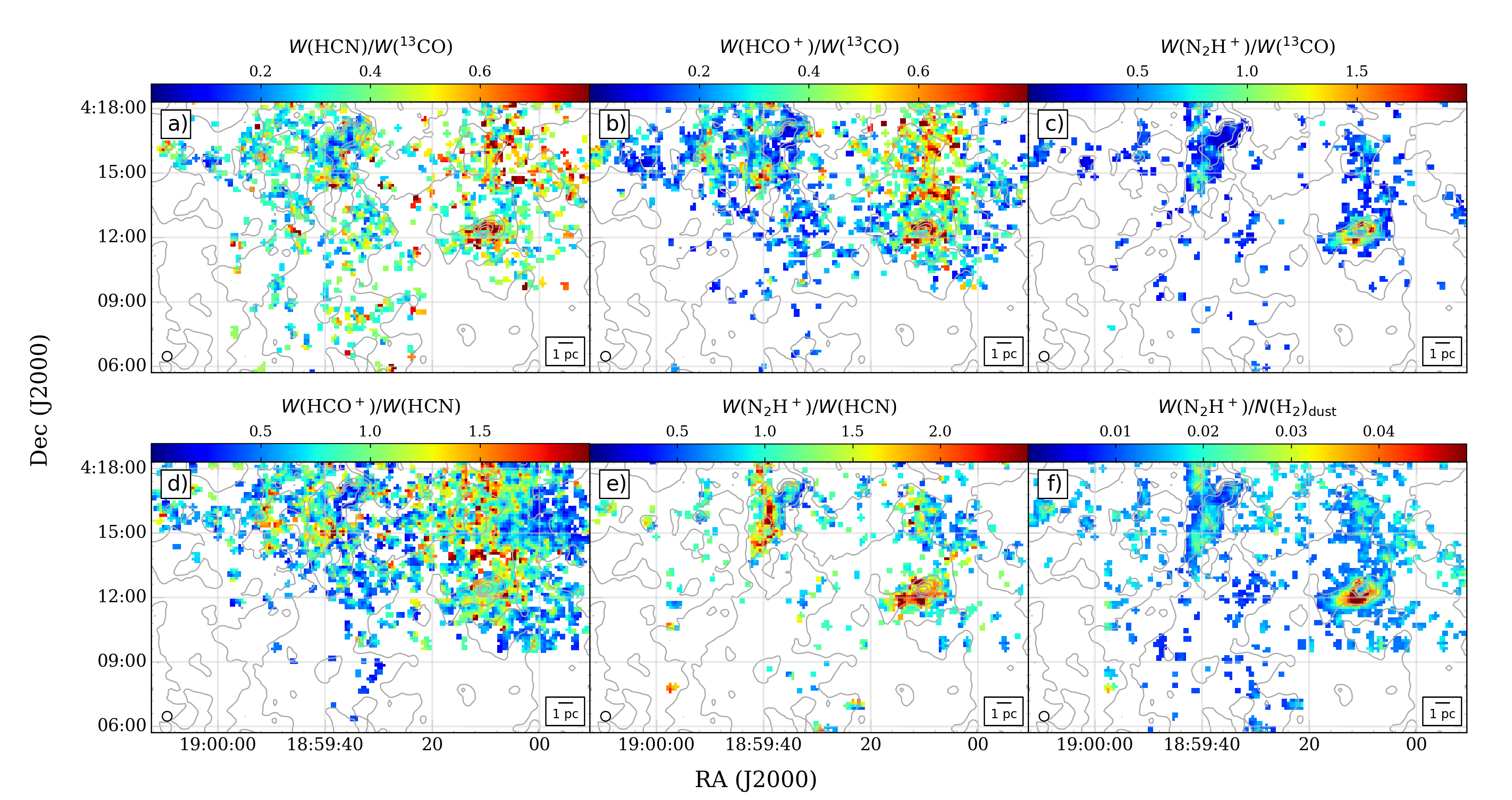}
    \caption{The same maps for Region\,4 as in Fig.~\ref{fig:reg1_ratio}. The gray contours mark the $N$(H$_2$)$_{\mathrm{dust}}$ H$_2$ column densities at the same levels as in Fig.~\ref{fig:reg4_morph}.}
    \label{fig:reg4_ratio}
\end{figure*}

\begin{figure*}
    \centering
    \includegraphics[width=\linewidth]{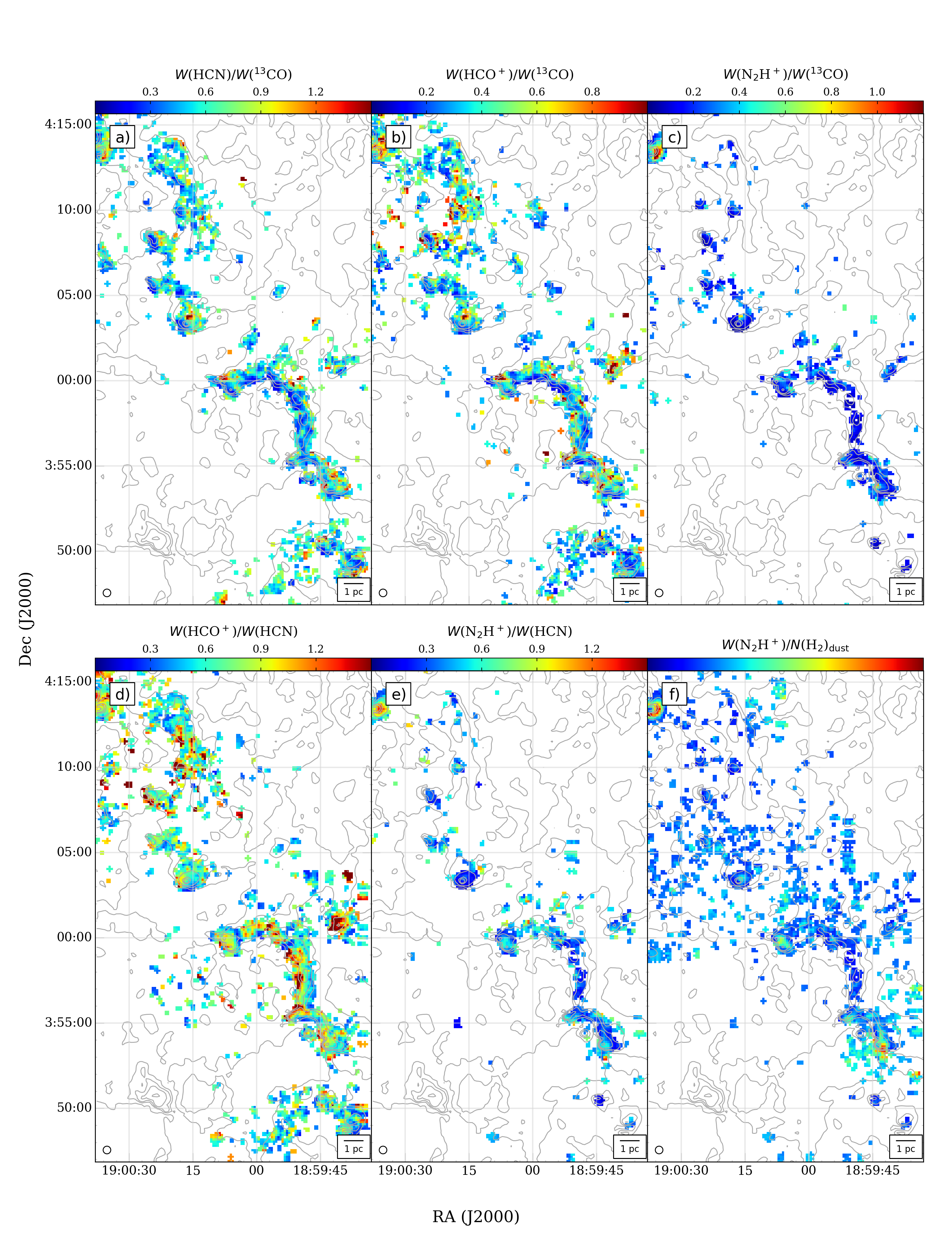}
    \caption{The same maps for Region\,5 as in Fig.~\ref{fig:reg1_ratio}. The gray contours mark the $N$(H$_2$)$_{\mathrm{dust}}$ H$_2$ column densities at the same levels as in Fig.~\ref{fig:reg5_morph}.}
    \label{fig:reg5_ratio}
\end{figure*}

\begin{figure*}
    \centering
    \includegraphics[width=\linewidth]{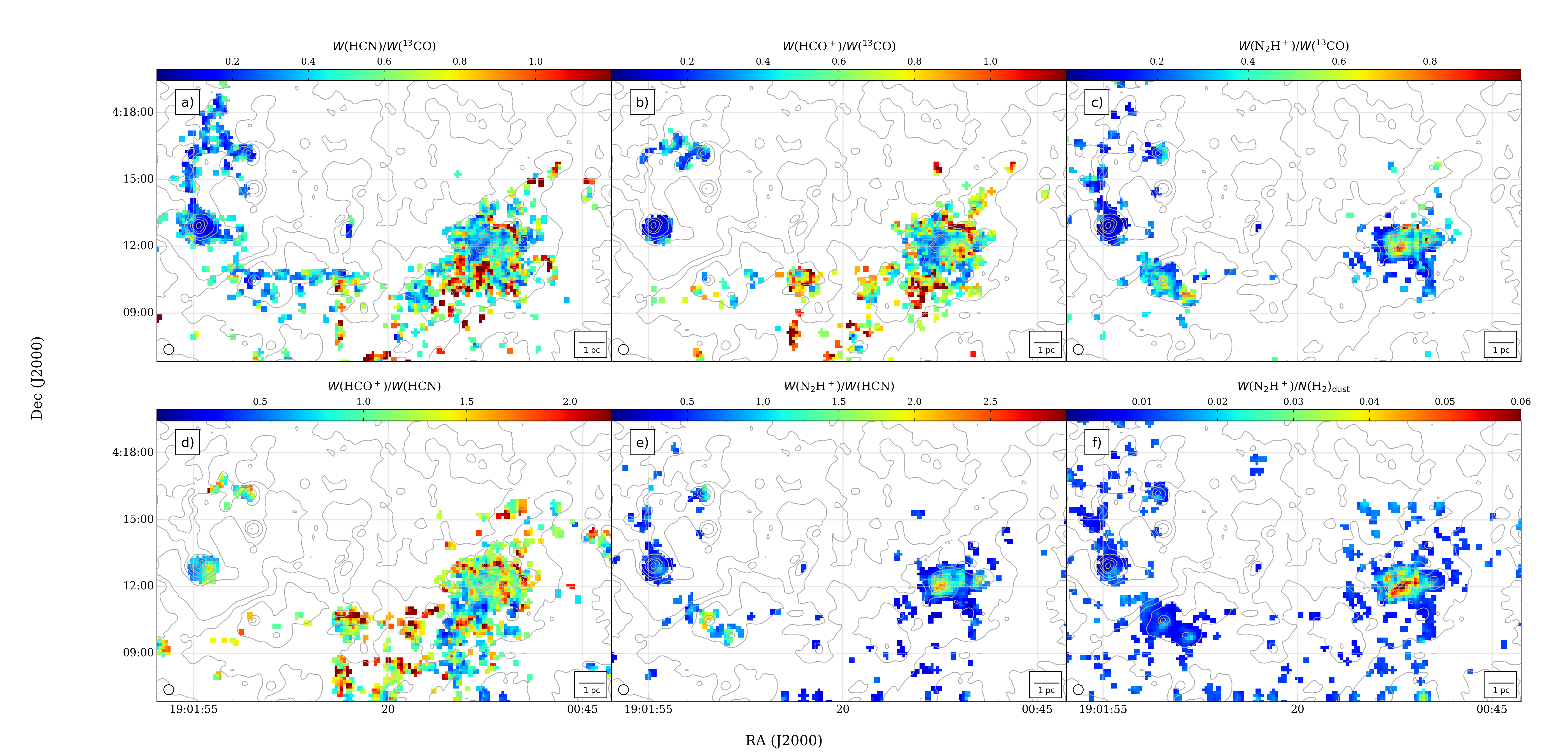}
    \caption{The same maps for Region\,6 as in Fig.~\ref{fig:reg1_ratio}. The gray contours mark the $N$(H$_2$)$_{\mathrm{dust}}$ H$_2$ column densities at the same levels as in Fig.~\ref{fig:reg6_morph}.}
    \label{fig:reg6_ratio}
\end{figure*}

\section{Clump average parameters}
\label{app:clumptable}

\begin{table*}
\resizebox{0.7\linewidth}{!}{
    \centering
    \begin{tabular}{l | rrrrrrr}
    \hline
    Object ID & \multicolumn{1}{c}{$W$($^{12}$CO)} & \multicolumn{1}{c}{$W$($^{13}$CO)} & \multicolumn{1}{c}{$W$(HCN)} & \multicolumn{1}{c}{$W$(HCO$^+$)} & \multicolumn{1}{c}{$W$(N$_2$H$^+$)} & \multicolumn{1}{c}{$N$(H$_2$)$_{\mathrm{dust}}$} & \multicolumn{1}{c}{$T_{\mathrm{dust}}$}  \\
          & \multicolumn{1}{c}{[K\,km\,s$^{-1}$]}  & \multicolumn{1}{c}{[K\,km\,s$^{-1}$]} & \multicolumn{1}{c}{[K\,km\,s$^{-1}$]} & \multicolumn{1}{c}{[K\,km\,s$^{-1}$]} & \multicolumn{1}{c}{[K\,km\,s$^{-1}$]} & \multicolumn{1}{c}{[10$^{22}$\,cm$^{-2}$]} & \multicolumn{1}{c}{[K]}  \\
    \hline
    \hline
    R1CL1   & 58.6 (4.0) & 21.1 (1.1) & 4.5 (0.2) & 3.8 (0.2) & 4.3 (0.3) & 1.96 (0.08) & 20.2 (0.08) \\
    R1CL2   & 92.6 (2.4) & 21.1 (1.4) & 3.2 (0.1) & 4.2 (0.2) & 2.5 (0.1) & 1.63 (0.02) & 20.2 (0.01) \\
    R1CL3   & 49.6 (2.5) & 8.4 (0.4)  & 2.2 (0.2) & 2.5 (0.1) & 1.8 (0.1) & 1.26 (0.01) & 19.8 (0.01) \\
    R1CL4   & 35.8 (2.1) & 5.2 (0.4)  & 1.5 (0.1) & 1.1 (0.0) & 1.7 (0.1) & 1.32 (0.01) & 19.4 (0.02) \\
    R1CL5   & 30.9 (2.5) & 4.6 (0.2)  & 2.6 (0.2) & 2.6 (0.1) & 2.6 (0.0) & 1.72 (0.03) & 18.7 (0.02) \\
    R1CL6   & 38.5 (1.6) & 6.2 (0.5)  & 1.6 (0.1) & 1.2 (0.1) & 2.6 (0.1) & 1.70 (0.03) & 19.7 (0.03) \\
    \hline
    Region\,1 clumps & 49.2 (18.3) & 12.0 (7.8)  & 3.1 (1.3) & 2.6 (1.3) & 2.9 (1.1) & 1.68 (0.29) & 19.8 (0.5) \\
    Region\,1 all      & 30.9 (12.4) & 6.6 (3.5) & 2.0 (0.8) & 1.8 (0.8) & 1.7 (0.8) & 1.04 (0.20) & 19.8 (0.3)  \\
    \hline
    R2CL1    & 65.1 (4.5) & 19.6 (0.8) & 4.0 (0.4) & 3.3 (0.2) & 2.4 (0.1) & 1.72 (0.09) & 20.9 (0.00) \\
    R2CL2    & 33.0 (1.5) & 6.3 (0.9)  & ...       & 3.0 (0.1) & 5.0 (0.2) & 1.24 (0.02) & 19.9 (0.03) \\
    R2CL3    & 45.2 (1.6) & 9.2 (0.9)  & 4.3 (0.1) & 2.6 (0.2) & 3.5 (0.0) & 1.93 (0.06) & 19.3 (0.05) \\
    R2CL4    & 36.5 (0.0) & 6.3 (0.0)  & 4.6 (0.0) & 2.2 (0.0) & ... & 1.57 (0.00) & 19.2 (0.00) \\
    \hline
    Region\,2 clumps & 42.4 (11.3) & 9.2 (4.7) & 4.3 (0.4) & 2.8 (0.4) & 4.1 (1.2) & 1.60 (0.32) & 19.8 (0.6) \\
    Region\,2 all      & 22.3 (9.0) & 5.7 (2.5) & 2.6 (0.9) & 2.1 (0.6) & 3.2 (0.9) & 1.00 (0.15) & 20.1 (0.2) \\
    \hline
    R3CL1  & 68.7 (3.7) & 18.5 (1.5) & 7.0 (0.4) & 5.8 (0.4) & 13.1 (0.8) & 3.78 (0.50) & 19.7 (0.24) \\
    R3CL2  & 26.5 (1.1) & 7.8 (0.3)  & 2.4 (0.1) & 1.5 (0.1) & 4.0 (0.4)  & 1.33 (0.02) & 19.5 (0.03)  \\
    R3CL3  & 30.4 (1.4) & 12.6 (0.9)  & ...       & 1.5 (0.1) & 2.1 (0.1)  & 1.78 (0.16) & 23.4 (0.33)  \\
    R3CL4  & 32.6 (0.9) & 15.9 (1.6) & 2.4 (0.1) & 1.5 (0.1) & 5.0 (0.2)  & 2.19 (0.17) & 20.7 (0.30)   \\
    R3CL5  & 49.8 (1.3) & 13.9 (1.2)  & 3.6 (0.2) & 2.8 (0.3) & 3.1 (0.3)  & 1.21 (0.04) & 20.0 (0.02)  \\
    \hline
    Region\,3 clumps  & 46.2 (23.2) &  15.7 (7.4) & 5.1 (2.8) & 3.6 (2.5) & 7.4 (5.4) & 2.38 (2.07) & 20.1 (1.4) \\
    Region\,3 all       & 19.7 (8.1) &  8.5 (3.3) & 2.4 (1.9) & 1.9 (1.0) & 2.2 (2.1) & 1.09 (0.23) & 20.2 (0.3) \\
    \hline
    R4CL1  & 37.8 (2.4) & 7.9 (0.3)  & 2.0 (0.3) & 2.8 (0.1) & 2.8 (0.4)  & 1.39 (0.02) & 20.2 (0.02) \\
    R4CL2  & 81.7 (4.3) & 18.3 (1.7) & 6.0 (0.4) & 5.9 (0.3) & 10.8 (0.6) & 3.74 (0.59) & 21.6 (0.32)  \\
    R4CL3  & 27.4 (1.0) & 24.5 (2.2) & 3.9 (0.3) & 1.6 (0.2) & 2.5 (0.1)  & 2.86 (0.13) & 20.8 (0.31)  \\
    R4CL4  & 32.6 (1.6) & 6.0 (0.0)  & 2.0 (0.0) & 1.6 (0.1) & 3.4 (0.1)  & 1.27 (0.01) & 20.3 (0.02)  \\
    R4CL5  & 34.6 (1.1) & 8.2 (0.9)  & 3.2 (0.5) & 2.8 (0.1) & 3.3 (0.2)  & 1.47 (0.01) & 19.9 (0.02)  \\
    R4CL6  & 34.4 (1.5) & 8.8 (0.9)  & 2.2 (0.1) & 2.8 (0.1) & 3.6 (0.1)  & 1.75 (0.02) & 20.1 (0.05) \\
    R4CL7  & 25.0 (0.9) & 7.2 (0.8)  & 2.0 (0.1) & 1.5 (0.1) & 2.9 (0.1)  & 1.34 (0.01) & 20.4 (0.10)   \\
    \hline
    Region\,4 clumps & 51.1 (32.0) &  14.8 (10.1) & 4.1 (2.6) & 3.7 (2.4) & 6.3 (4.7) & 2.48 (2.74) & 20.8 (1.6) \\
    Region\,4 all      & 23.1 (8.8) &  7.5 (3.2) & 2.1 (0.9) & 1.6 (0.8) & 2.6 (2.0) & 1.25 (0.35) & 20.3 (0.7) \\
    \hline
    Interarm clumps & 48.6 (26.5) & 14.1 (8.7) & 4.2 (2.5) & 3.4 (2.2) & 6.0 (4.8) & 2.27 (2.22) & 20.3 (1.4) \\
    Interarm all      & 23.2 (10.1) & 7.1 (3.3) & 2.2 (0.9) & 1.7 (0.9) & 2.4 (1.9) & 1.13 (0.29) & 20.2 (0.5)  \\
    \hline
    \hline
    R5CL1 & 125.1 (6.5) & 31.5 (2.1) &  8.8 (0.7)  & 7.8 (0.5) & 5.0 (0.4) & 2.76 (0.11) & 20.3 (0.26) \\
    R5CL2 & 73.0 (0.0) & 27.1 (0.0) &  8.3 (0.0)  & 4.8 (0.0) & 1.3 (0.0) & 2.26 (0.00) & 20.6 (0.00)  \\
    R5CL3 & 68.7 (11.4) & 23.8 (3.0) &  5.8 (0.0)  & 5.7 (0.4) & 4.2 (0.1) & 2.35 (0.29) & 20.1 (0.15)  \\
    R5CL4 & 95.6 (7.3) & 33.6 (4.0) & 10.0 (0.8)  & 8.7 (0.4) & 5.4 (0.6) & 3.17 (0.24) & 23.4 (0.37)  \\
    R5CL5 & 79.2 (8.0) & 25.3 (3.4)  & 4.6 (0.2)   & 3.6 (0.2) & 2.6 (0.2) & 1.94 (0.09) & 20.3 (0.03)  \\
    R5CL6 & 141.0 (10.6) & 58.2 (6.7) & 19.3 (1.0)  & 10.2 (0.6)& 5.5 (0.5) & 4.27 (0.49) & 24.5 (0.31) \\
    R5CL7 & 59.4 (5.4) & 14.8 (0.9)  & 2.7 (0.2)   & 5.1 (0.0) & 1.7 (0.0) & 2.12 (0.06) & 21.0 (0.00)  \\
    R5CL8 & 107.3 (7.2) & 31.8 (4.2) & 12.8 (0.6)  & 10.0 (0.5)& 13.8 (1.0)& 7.00 (1.10) & 19.5 (0.21)  \\
    R5CL9 & 58.1 (5.2) & 33.8 (2.5) & 6.8 (0.5)   & 3.4 (0.5) & 2.4 (0.2) & 2.23 (0.17) & 21.5 (0.12)  \\
    \hline
    Region\,5 clumps & 104.9 (34.8) & 34.2 (17.4) & 10.9 (4.8) & 8.3 (2.7) & 7.2 (5.0) & 4.14 (3.27) & 21.4 (2.1) \\
    Region\,5 all      & 23.5 (18.6) & 10.4 (7.5) & 3.4 (2.1) & 2.1 (1.5) & 1.8 (1.7) & 1.27 (0.41) & 20.7 (0.7) \\
    \hline
    R6CL1 & 112.6 (2.0) &  43.9 (2.1) & 12.1 (0.3) & 12.9 (0.4) & 18.8 (1.0) & 5.50 (0.52) & 21.5 (0.11)   \\
    R6CL2 & 86.6 (5.6) & 33.3 (4.0) & 4.8 (0.6)  & 6.0 (0.7)  & 3.8 (0.3)  & 3.76 (0.73) & 19.5 (0.32)   \\
    R6CL3 & 52.2 (1.9) & 11.4 (0.3)  & ...        & 3.5 (0.0)  & 6.0 (0.2)  & 3.83 (0.24) & 18.7 (0.14)   \\
    R6CL4 & 57.2 (3.2) & 17.9 (1.2) & 2.3 (0.0)  & ...        & 6.5 (0.4)  & 4.91 (0.47) & 19.4 (0.11)   \\
    R6CL5 & 72.2 (0.6) & 23.3 (2.9) & 6.6 (1.0)  & 8.0 (0.2)  & 6.1 (0.5)  & 4.09 (0.55) & 20.7 (0.43)   \\
    R6CL6 & 58.8 (1.9) & 14.0 (1.0)  & 3.1 (0.2)  & ...        & 1.8 (0.1)  & 2.03 (0.06) & 20.9 (0.08)   \\
    R6CL7 & 15.3 (0.9) & 115.2 (9.7) & 3.6 (0.3)  & 3.5 (0.0)  & 5.8 (0.3)  & 6.05 (0.99) & 27.4 (0.40)   \\
    \hline
    Region\,6 clumps & 73.0 (31.5) &  36.6 (31.9) & 8.0 (4.4) & 10.8 (3.4) & 9.8 (7.4) & 4.26 (2.45) & 21.3 (2.3) \\
    Region\,6 all      & 28.2 (18.1) &  12.4 (11.2) & 4.4 (2.2) & 5.1 (2.2) & 3.4 (3.7) & 1.41 (0.49) & 20.2 (0.6) \\
    \hline
    Arm clumps & 86.3 (36.5) & 35.6 (26.7) & 9.5 (4.8) & 9.3 (3.2) & 8.6 (6.6) & 4.21 (2.83) & 21.3 (2.2)   \\
    Arm all      & 25.6 (18.5) & 11.2 (9.3) & 3.8 (2.2) & 2.8 (2.1) & 2.4 (2.7) & 1.33 (0.45) & 20.5 (0.7)  \\
    \hline
    \end{tabular}
    }
    \caption{Average parameters of the N$_2$H$^+$-clumps, observing regions, and the two GMFs. (1) Object name; (2, 3, 4, 5, 6) average integrated intensities of the studied molecular species in the object; (7) average column density; (8) average dust temperature. The clump, region and GMF molecular emission averages were all computed by only taking pixels above 2\,$\times$\,$\sigma_{\mathrm{int}}$. The errors listed for the clumps are the error of the mean, while the errors for the observing regions and the filaments are the standard deviation.}
    \label{tab:clusters}
\end{table*}


\bsp	
\label{lastpage}
\end{document}